\documentclass[12pt,preprint]{aastex}
\usepackage{framed} 
\newcommand{\be}{\begin{equation}}
\newcommand{\ee}{\end{equation}}

\def\lta{\,\raise 0.3 ex\hbox{$ < $}\kern -0.75 em
 \lower 0.7 ex\hbox{$\sim$}\,}
\def\gta{\,\raise 0.3 ex\hbox{$ > $}\kern -0.75 em
 \lower 0.7 ex\hbox{$\sim$}\,}

\newcommand{\swig}{ {\tilde s}}
\newcommand{\weber}{ {W}_{\scriptstyle\rm EB}}
\newcommand{\surf}{ \zeta_{\rm s}}
\newcommand{\tauop}{ \tau_{\rm op}}
\newcommand{\fdrop}{ f_{\rm d} }
\newcommand{\fstar}{f_\ast}
\newcommand{\smin}{{\tilde s}_{\rm min}}
\newcommand{\width}{\sigma_{\rm w}}

\begin{document} 

\title{\bf Production of White Dwarf Debris Disks\\
from Planetary Collisions and Poynting-Robertson Drag} 

\author{Fred C. Adams$^{a,b,}$\footnote{Corresponding author:
    Fred C. Adams, fca@umich.edu} $\,$ and Konstantin Batygin$^c$}

\affil{$^a$Physics Department, University of Michigan, Ann Arbor, MI
  48109, USA}

\affil{$^b$Astronomy Department, University of Michigan, Ann Arbor, MI
  48109, USA}

\affil{$^c$Division of Geological and Planetary Sciences, California
  Institute of Technology, Pasadena, CA 91125, USA} 

\begin{abstract}
{This paper explores the production of debris disks around white
dwarfs, which are observed to experience frequent pollution events.
In this scenario, planetary collisions occur just outside the surface
of a highly evolved AGB star, after the end of its mass loss phase, as
it starts its transition into a white dwarf. The collisions vaporize a
fraction of the planetary material, which subsequently expands, cools,
nucleates droplets, and condenses into rocky bodies with
characteristic size of order $s\sim1$ cm. The rocky debris disk
evolves through the action of Poynting-Robertson drag, driven by the
stellar luminosity, which is initially large but steadily decreases
with time. This scenario provides an alternate channel for delivering
rocky material to the immediate vicinity of a white dwarf. }
\end{abstract}

\section{Introduction}
\label{sec:intro} 

A significant fraction of observed white dwarfs exhibit atmospheric
pollution by heavy elements 
\citep{farihi2016,xu2024,veras2024,malamud2026},
despite the fact that metals should rapidly settle out of their
atmospheres under the strong surface gravity of the star.  The
observed persistence of photospheric metals thus requires ongoing
and/or recent accretion of external material with high metallicity.
Infrared excesses, gaseous emission features, and detailed abundance
measurements jointly indicate that the accreted material is
predominantly rocky and broadly similar in composition to terrestrial
planets. In addition to their intrinsic interest, these polluted white
dwarfs and their associated debris disks provide an important probe of
the long-term evolution and composition of planetary systems
{(see also \citealt{mustill2026})}. 

The current picture of white dwarf pollution emerged through a series
of theoretical and observational advances over the past two decades.
Early work suggested that remnant planetary systems could survive
post-main-sequence evolution and then perturb asteroids or minor
planets onto star-grazing orbits after the host star becomes a white
dwarf \citep{debes2002,jura2003,zuckerman2003,frewen2014}. Subsequent infrared
observations detected compact dusty disks around a growing number of
polluted white dwarfs \citep{reach2005,kilic2006,jura2007,farihi2009},
although the frequency of disk detections is only $\sim2\%$ (see Table
1 of \citealt{farihi2016}). {These disks typically show infrared
excesses out to $\sim24\mu$m, and are thus relatively compact, which
provides important observational constraints on pollution scenarios.}
In addition, spectroscopic studies demonstrated that the atmospheric
abundances are consistent with accretion of rocky, but volatile-poor
material, i.e., the composition is more like terrestrial planets than
interstellar dust grains
\citep{zuckerman2007,klein2010,zuckerman2010,jura2014,xu2019,blouin2022}.
This body of work provides compelling evidence that planetary material
survives stellar evolution and continues to dynamically interact over
Gyr timescales (see the reviews of \citealt{farihi2016};
\citealt{xu2024,veras2024,malamud2026,mustill2026};
and references therein).

In the currently favored scenario, rocky bodies orbiting at distances
of order $\sim1$ AU are destabilized after stellar mass loss and are
scattered into highly eccentric orbits that pass within the tidal
disruption radius of the white dwarf. Numerical studies have shown
that such objects are tidally disrupted into eccentric streams or
rings of debris, which subsequently evolve into compact disks through
collisional and radiative processes \citep{veras2014a,veras2020}.
{Some observed WDs have gaseous disks that are moderately
eccentric \citep{manser2016}, consistent with an origin from
dynamically delivered parent bodies}.  Dynamical pathways capable of
driving bodies into the Roche limit of the white dwarf include
planet-planet scattering, secular resonances, Lidov-Kozai forcing in
binaries, and delayed instabilities triggered by stellar mass loss
\citep{bonsor2011,debes2012,veras2014a,hamers2016,pichierri2018}.
{Additional sources of metal-rich debris include ejecta from 
impacts \citep{veraskuros} and driving asteroids to break-up
conditions through radiative effects \citep{veras2014b,veras2022}.}

{Related observations \citep{vanderburg2015} provide evidence
for transits of comet-like bodies in some white dwarf systems. This 
finding demonstrates that large objects can be transported to the
vicinity of the stars. The occurrence rate is highly uncertain, but
could be large enough to account for the observed levels of WD
pollution \citep{robert2024}. In addition, some objects reside in wide
orbits \citep{vanderbosch2020}, although the eccentricities are not
known. }

Despite the substantial successes of this existing framework, several
important issues remain unresolved. One difficulty concerns the
efficiency and timing of material delivery into the immediate vicinity
of the white dwarf. In many models, the parent bodies originate at
distances comparable to or larger than several AU and must survive
both giant-branch evolution and subsequent dynamical transport.
Another challenge is the production of sufficient quantities of rocky
bodies and dust. Given the observed accretion rates
${\dot M} \approx 10^6-10^{10}$ g s$^{-1}$ (see Fig. 2 of
\citealt{farihi2016}), and lifetimes of order 1 Gyr, the required
mass supply falls in the range $\Delta M=\int{\dot M}dt \sim$
$3\times10^{22}$ to $3\times10^{26}$ g, up to the mass of a dwarf
planet. Finally, numerical calculations of tidal disruption often
produce highly eccentric streams or collisionless rings that require
additional mechanisms to circularize and generate the compact dusty
disks inferred observationally.

\begin{figure}
\centering
\includegraphics[width=1.0\linewidth]{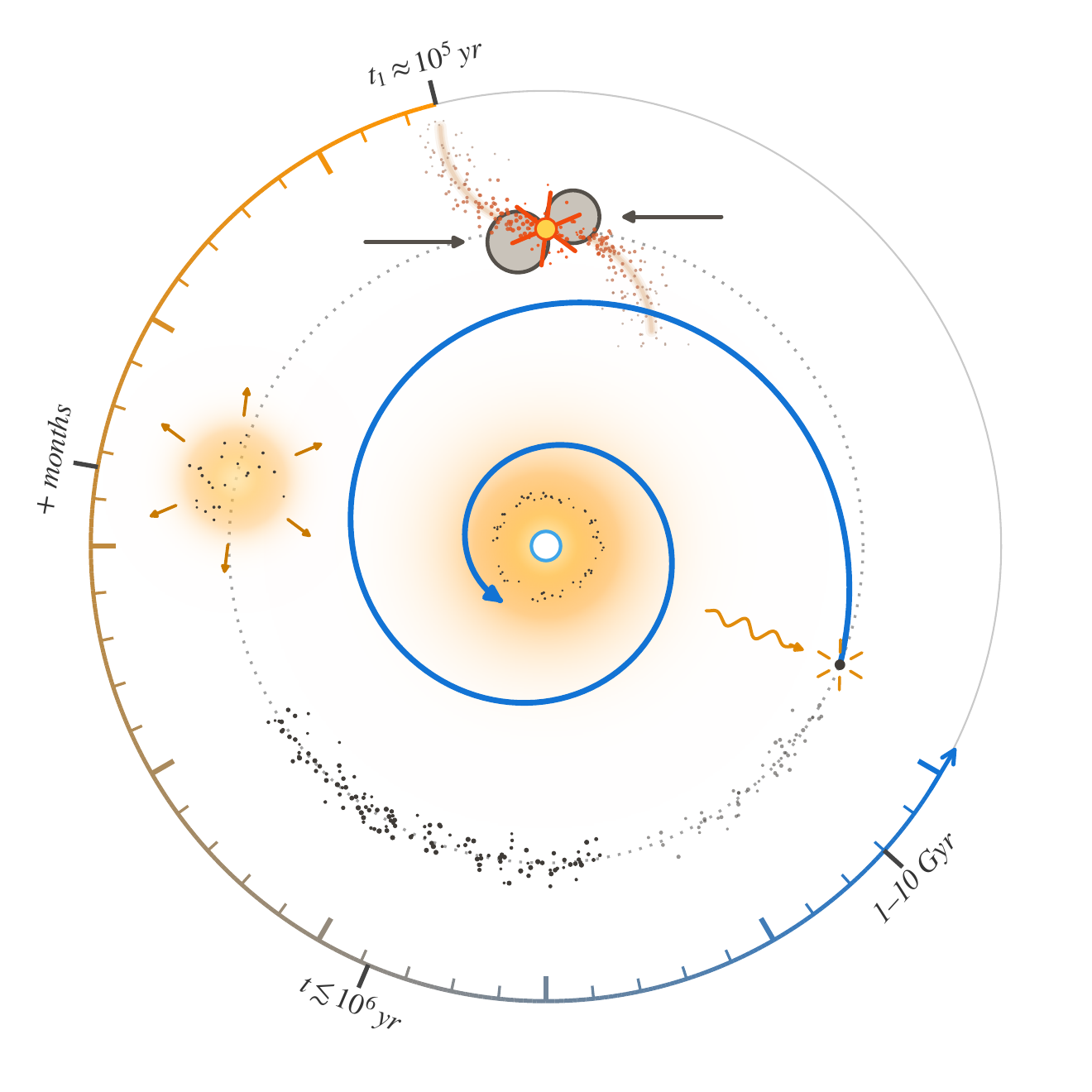}
\vskip-0.10truein
\caption{Schematic depiction of this paradigm for debris disk
formation around white dwarfs. After stellar mass is lost during the
AGB phase, planets collide and vaporize a fraction of their mass,
thereby producing an expanding gas cloud in the center of mass frame,
with the planetary remnants leaving the vicinity. The expanding vapor
cools and nucleates liquid droplets, which condense to produce rocky
debris. As the star contracts to become a white dwarf, the rocks move
inward, driven by PR drag with time-dependent stellar luminosity. The
end result is a debris disk that continues to supply rocky material to
the white dwarf for billions of years. } 
\label{fig:schematic}
\end{figure}

This paper constructs an alternative and potentially complementary
pathway for generating white dwarf debris disks (see Figure
\ref{fig:schematic}). Rather than beginning with the tidal disruption
of distant minor bodies after inward scattering and white dwarf
formation, we consider the case where rocky planets or somewhat
smaller bodies collide after --- or perhaps during --- the asymptotic
giant branch (AGB) phase itself, at orbital locations just exterior to
the extended stellar envelope. At this evolutionary stage, orbital
expansion driven by stellar mass loss leads to dynamical instabilities
among surviving planets, thereby naturally enhancing the probability
of close encounters and high-velocity impacts. In particular, previous
work shows that planetary systems that are stable during the
main-sequence phase do not need to be close-packed in order to become
unstable during the phases of mass loss \citep{veras2013,mustill2018}.
The exact mode of instability depends on the specific dynamical
architecture during mass loss (compare \citealt{pu2015,zink2020}; and
others). In any case, planetary systems often reside close to marginal
stability \citep{laskar1996}, so that modest changes to their underlying
parameters are likely to instigate instability. As a result, this
paper starts with the assumption that planetary collisions occur.

Collisions occurring near the surface of an AGB star have favorable
parameters for producing the rocky debris of interest. The orbital
velocities near the giant star (distances of a few AU) are typically
$\sim20-40$ km/s during the AGB phase and decrease to $\sim10-20$ km/s
after mass loss. For comparison, the escape speeds for earth-like and
super-earth planets fall in the range $\sim10-20$ km/s. As a result, both
planetary collisions and ejections are expected to take place. Since
about half of the white dwarfs show observational signatures of
pollution, a branching ratio of $\sim1/2$ for collisions vs ejections
is consistent with current data. When collisions occur, the planetary
escape speeds lead to collision speeds that readily exceed the
vaporization threshold for silicate material.  Impacts are thus
likely, and a substantial fraction (typically $f\sim10\%$ -- see
Section \ref{sec:collide}) of the rocky material involved in the
collision can be shock-heated and vaporized.  The expanding vapor
subsequently cools and condenses into droplets, and then solid
fragments, analogous to the silicate condensates produced in other
high-temperature astrophysical environments (e.g., chondules;
\citealt{asphaug2011,melosh1991}). The characteristic size
distribution of the condensed material is determined by the
competition between surface tension, turbulent fragmentation,
coagulation, and aerodynamic stresses in the expanding flow
{(e.g., see \citealt{benz2007}).} 

Once formed, the condensed droplets evolve within the complicated
circumstellar environment produced after late AGB evolution. Depending
on their sizes, and the timing of the events, the particles can
experience outward forces due to the stellar wind and stellar
radiation pressure, as well as inward torques driven by gas drag and
the Poynting-Robertson (PR) effect. These interactions are highly time
dependent, as the star transitions from the AGB phase to the white
dwarf stage. In this picture, the white dwarf debris disk is not
produced primarily by the late tidal disruption of a single asteroid,
but instead represents the surviving remnants of a collisional cascade
(and vaporization) initiated during the final stages of stellar
evolution. The resulting rocky remnants are then transported inward
due to PR drag. The collisions must take place while the luminosity of
the star is still large (compared to a typical white dwarf), but after
most of the stellar mass loss has occurred. {One feature of this
scenario is that} PR drag moves the rocky remnants larger distances,
driven by the large but decreasing luminosity of the star
as it cools from its AGB phase to the white dwarf phase {(see
also \citealt{veras2015,veras2022} for related work with
$dL_\ast/dt\ne0$).} In contrast, {most} previous work has
considered PR drag acting in the compact debris disks that surround
white dwarfs at later stages with constant (smaller) luminosity (see
\citealt{rafikov2011,bochkarev2011}). {For completeness, note
that white dwarf surface temperatures are higher than silicate
sublimation temperatures, so that the final accretion onto the star
takes place while the material is in gaseous form.}

This scenario offers several potential advantages. First, the rocky
source material naturally moves inward to orbital radii comparable to
the eventual dimensions of observed white dwarf debris disks. Second,
the impact process directly produces large quantities of small
condensed solids, and thus circumvents some of the difficulties
associated with converting eccentric tidal streams into dusty disks.
Third, the total mass reservoir can be substantial if the collisions
involve super-Earths or similarly large rocky bodies (even though only
a fraction of the total mass is vaporized).  Fourth, the material
produced through this mechanism has the composition of rocky planets,
consistent with observations. Moreover, because the material is
processed through a high-temperature vapor phase, the resulting
condensates may exhibit distinctive compositional or structural
signatures relevant to the observed abundances in polluted white
dwarfs \citep{xu2021}. 

The goal of this work is to develop this scenario for producing rocky
debris around white dwarfs. The paper is organized as follows.
Planetary collisions and rock production are considered in Section
\ref{sec:collide}. This discussion includes the energetics of
planetary collisions near AGB stars, an expansion model for the
resulting high-metallicity gas, the condensation of the vapor into
droplets and solid fragments, and the characteristic droplet sizes
produced by condensation and coagulation in the expanding flow. The
dynamical evolution of the condensed solids is then considered in
Section \ref{sec:prdrag}, which shows how drag forces due to the
Poynting-Robertson effect with time dependent stellar luminosity can
move rocky debris inward. The global evolution of the disk of rocky
material is then considered in Section \ref{sec:evolve}. The paper
concludes, in Section \ref{sec:conclude}, with a summary of our
results and a discussion of their implications. This paper shows that
planetary collisions near the surfaces of AGB stars can plausibly
generate long-lived reservoirs of rocky material that evolve into
debris disks around white dwarfs.

\section{Production of Liquid Silicate Droplets and Rocks
  from Planet Collisions}
\label{sec:collide}

During and after the AGB phase of stellar evolution, as the star loses
mass and planets adjust their orbits accordingly, the changes in
system architecture often result in chaos, instability, and planetary
collisions (e.g., \citealt{debes2002,veras2013,veras2016,mustill2018,
maldonado2020}; {\citealt{maldonado2021}}).  {The timing of
the collisions depends on a number of factors, including the system
architecture, the planet masses, and the amount of stellar mass loss
(see also \citealt{verasetal2016}).  Both the collision rate and the
ejection rate tend to peak just after the phase of stellar mass loss
--- for the standard case of steady outflow --- but disruption
continues over much longer time scales of 100 Myr to even 1 Gyr.
However, recent work suggests that stellar outflows are episodic
and asymmetric, which explains the observed WD kinematics
\citep{fuller2026}, and which implies more rapid disruption of
planetary sytems (\citealt{batygin2026}; see also Section
\ref{sec:conclude}). } This section begins with the assumption that
planetary collisions occur, and argues that they result in the
vaporization of a significant portion of rocky material.\footnote{As
shown in subsequent sections, the amount of rocky material
delivered to the white dwarf decreases for collisions that occur at
larger WD ages.}  The present discussion thus starts where the
aforementioned previous work ends.  The subsequent evolution of this
expanding gas cloud results in the production of droplets, and then
rocky condensations, as shown here. We assume that the collisions
vaporize a substantial amount of mass $M$, which is expected to be
roughly comparable to that of Earth.

\subsection{Collision Requirements}
\label{sec:collison} 

The collision must be sufficiently energetic to both unbind the
planets and vaporize the planetary material. These requirements
can be written in the form  
\be
{1\over2} M v_{\rm col}^2 \gta \alpha {GM^2\over R_{\rm p}} \qquad
\Rightarrow \qquad v_{\rm col}^2 \gta {GM\over R_{\rm p}} \,,
\ee
and
\be
{1\over2} M v_{\rm col}^2 \gta M C (\Delta T) \qquad \Rightarrow
\qquad v_{\rm col}^2 \gta 2 C (\Delta T) \,,
\ee
where $C$ is the specific heat and $(\Delta T)$ is the necessary
temperature increment. The first constraint is that the collision
speed must be comparable to the escape speed (which should be 10 to 20
km/s) in order to break apart the planet.  For the second constraint,
if we use $C=7\times10^6$ erg/g/K and
$(\Delta T) = T_{\rm melt}\approx$ 1700 K, then one requires
$v_{\rm col} \gta 1$ km/s for melting.  Using the temperature
increment $(\Delta T) = T_{\rm vapor} \approx 4000$ K, then one
requires a minimum collision speed $v_{\rm col} \gta 2$ km/s for
vaporization. The energy required to break apart the planet is thus
significantly greater than the energy required for vaporization of
the rock.

Since the planets collide by falling toward each other through their
gravitational potential wells, the collision velocity is of order
the escape speed. More specifically, working in the center of mass
(CM) frame for equal mass planets, the speed at impact is given by 
\be
v_{\rm col}^2 = {1\over4} v_{\rm esc}^2 + v_\infty^2 \,, 
\ee
where $v_\infty$ is the speed of the bodies (in the CM frame) in the
absence of gravitational acceleration. This speed is determined by the
trajectories of the planets in their Keplerian orbits, and is expected
to be of order $v_\infty\sim{e}\,v_{\rm orb}$, where $e$ is the
eccentricity.

\begin{figure}
\centering
\hspace*{-2cm}\includegraphics[width=0.80\linewidth]{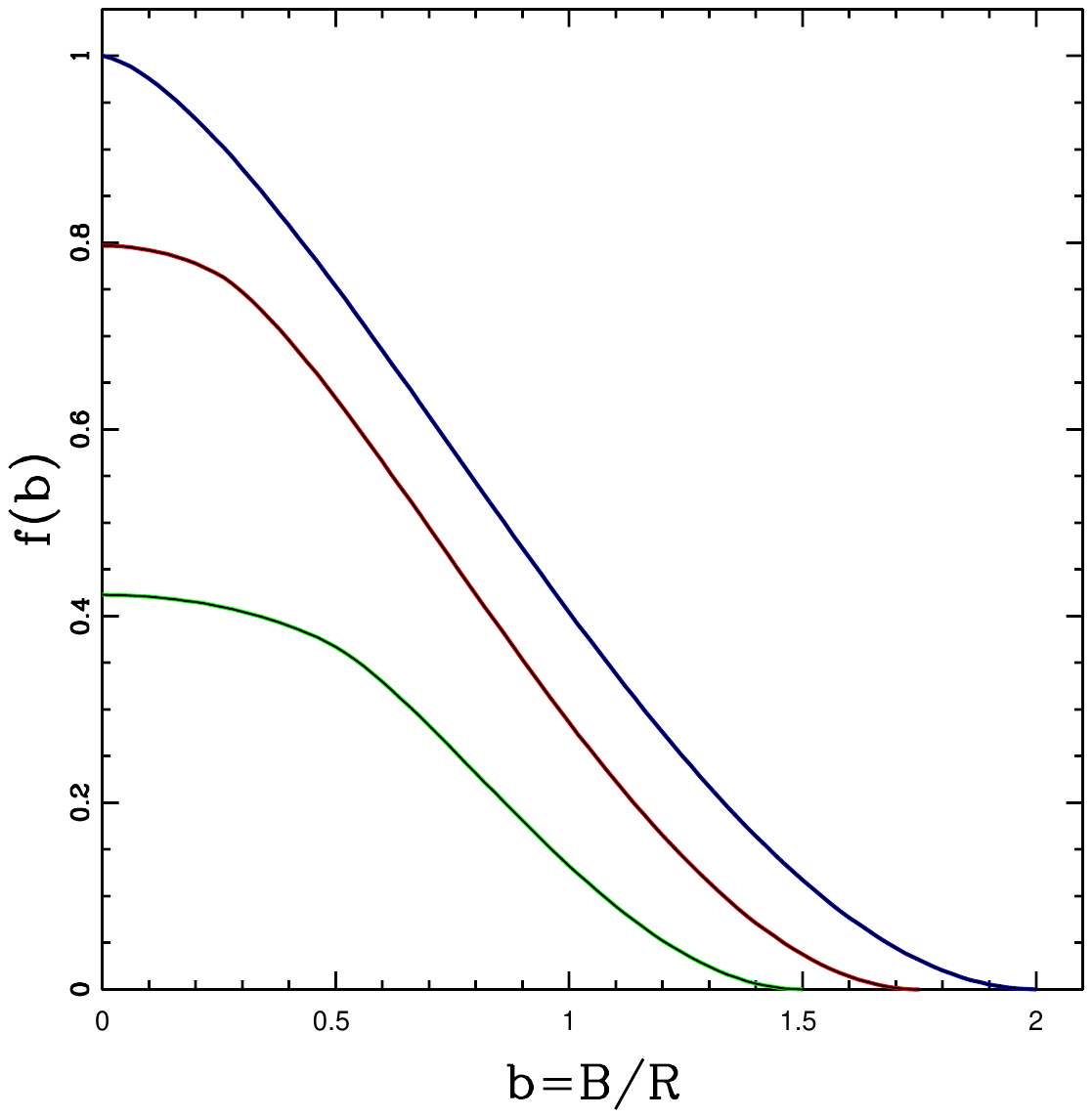}
\vskip-1.0truein
\caption{Fraction of the volume of two colliding planets that
experience impact. The fraction is plotted as a function of impact
parameter, defined to be the distance between planet centers at the
point of closest approach. The upper blue curve shows the fraction for 
equal sized planets. Results are also shown for cases where the
smaller planet has radius $R_{\rm small}/R_{\rm large}$ = 0.75 (red curve)
and $R_{\rm small}/R_{\rm large}$ = 0.5 (green curve). }
\label{fig:fraction}
\end{figure}

In general, the planets do not collide head-on, but rather at an angle
$\theta$, or equivalently with non-zero impact parameter $B$.  For
equal mass planets of radius $R$, the impact parameter $B$ =
$2R\sin\theta$ (e.g., \citealt{leinhardt2012}). The most likely
collision angle $\theta=\pi/4$ \citep{shoemaker1962} corresponds to
the impact parameter $B=\sqrt{2}R$. For $B\ne0$, only a fraction of
the material in the colliding planets will experience an impact. Here
we define $f(b)$, where $b=B/R$, to be the fraction of the volume of
the planets that experiences overlap as the two bodies pass by each
other. Consider the first planet to be at rest, while the second
planet passes by at impact parameter $B$. The second planet sweeps out
a cylinder of radius $R$ that is displaced from the origin by distance
$B$. The fraction $f$ is thus the overlap integral between the first
planet (a sphere of radius $R$ centered at the origin) and the swept
volume of the second planet (an infinite cylinder of radius $R$
displaced from the origin by a distance $B$).  The result (from
numerical evaluation) is shown in Figure \ref{fig:fraction} for
planetary collisions with equal and unequal masses (sizes). 

The picture that emerges can be described as follows: When the two
planets collide with nonzero impact parameter, only part of the total
volumes of the two colliding planets interact with each other. As a
result, only the fraction $f(b)$ will collide, shock, and vaporize.
The remaining pieces of the planets -- the top part of one and the
bottom part of the other -- will continue on their trajectories and
leave the scene of the collision. The colliding fraction will remain
at rest with net zero momentum in the CM frame. The key feature of
this type of collision is that the impact has collision velocity given
by the full masses of the planets. But since only a fraction of the
planet mass is involved in the interaction, the remaining vaporized
gas cloud will have much less self-gravity, and will not be bound. 

We can write down the energy budget as follows: The colliding portion
of the planet has initial kinetic energy $M_{\rm int}v_{\rm col}^2/2$
and self-gravity $-\alpha GM_{\rm int}^2/R_{\rm int}$, where the
interaction mass $M_{\rm int} = 2f(b)M$. The energy required for
vaporization is given by $E_{\rm vap} = M_{\rm int} C ({\Delta T})$
$\equiv M_{\rm int} v_{\rm vap}^2$, where we have defined
$v_{\rm vap}^2 \equiv C (\Delta T)_{\rm vap} \sim$ (1.7 km/s)$^2$.
After the collision, conservation of energy implies that the energy
$E=M_{\rm int}v_{\rm flow}^2/2$ remaining in the expanding gas cloud
is given by 
\be
v_{\rm flow}^2 = {1\over4} (1-2f[b]) v_{\rm esc}^2 + v_\infty^2 - 2
v_{\rm vap}^2\,.
\label{vflow} 
\ee
Here we have assumed that $\alpha=1/2$ and $R_{\rm int}=2R$ to
evaluate the self-gravity term. For the benchmark impact parameter
where $b=\sqrt{2}$, the interaction fraction $f(b)\approx0.157$. The
final two terms are comparable in magnitude, and $(1-2f)\sim0.7$, so
that the specific energy of the expanding gas cloud is large enough to
produce a substantial flow velocity. For example, in the limit
$v_\infty\to0$ with $f=0.2$, $v_{\rm esc}=20$ km/s, and
$v_{\rm vap}=2$ km/s, one finds $v_{\rm flow}\approx$ 7.2 km/s.

\subsection{Expansion of the Gaseous Cloud}
\label{sec:expansion} 

After the collision, some fraction of the mass ends up in the form of
vapor with the composition of the original rock.  For the sake of
simplicity, we model this structure as a spherically expanding gas
cloud. In addition, any remaining planetary remnants are assumed to
have different velocities, so that they separate from the cloud. As it
expands, the gas cloud cools, and will eventually attain an expansion
speed $v$. As outlined above, this speed must be comparable to but
smaller than both the escape speed (about 15 km/s) and the orbital
speed (about 30 km/s), so the canonical value $v\sim3-10$ km/s. Just
after the collision, the gas must be accelerated by $PdV$ work up
attain these speeds.  Relatively soon thereafter, the gas will expand
homologously, with the size and density given below.

To leading order, the expansion speed is constant at later times,
so that the size of the expanding gas cloud is given by 
\be
R(t) = vt \,.
\ee
Note that the late-time expansion speed $v$ is somewhat smaller than
the flow speed estimated from equation (\ref{vflow}), as energy is
dissipated in order for the flow to reach its constant velocity state.
At this epoch, the gas density takes the form 
\be
\rho_{\rm g} = {3M\over4\pi R^3} = {3M\over4\pi v^3 t^3} \,.
\label{density} 
\ee

Next we consider the thermodynamics of the expansion. The gas will
generally expand adiabatically, so that the first law of
thermodynamics reads 
\be
dU + PdV = 0 \,.
\ee
For an ideal gas, this expression can be written in the form
\be
{1 \over \gamma - 1} {dT \over T} + {dV \over V} =
{1 \over \gamma - 1} {dT \over T} + 3{dR \over R} = 0\,,
\ee
which integrates to become
\be
T R^{3(\gamma-1)} = {\it constant} \qquad \Rightarrow \qquad
T \sim R^{-3(\gamma-1)} \,. 
\ee
So we find $T \sim R^{-1}$ for $\gamma=4/3$ (and $T\sim R^{-2}$ for
$\gamma=5/3$). The interpretation is that the gas will cool rapidly as
it expands, with little remaining thermal energy. The random motions
of the gas will be dominated by the turbulent motions, which will take
longer to dissipate. Using parameter values from the previous
section, the starting temperature is $\sim5-15$ times higher than the
vaporization temperature, so that the cloud must expand by the same
factor to become cool enough for condensation (for $\gamma=4/3$).
This radius (roughly $R_{\rm cool}\sim0.001$ AU) is much smaller than
the Hill radius ($R_H\sim0.1$ AU; see Section \ref{sec:boundary}).

{More quantitatively, the saturation vapor pressure for silicates
can be written in the form
\be
\log_{10} \left( {P_{\rm sat}\over1\,{\rm bar}}\right) = - A
- {T_c\over T} + B \log_{10} \left( {T\over1\,{\rm K}} \right)\,,
\ee
where $T$ is the tempeature and the constants have fitted values
$A\approx30.6757$, $T_c\approx8228.146$ K, and $B\approx$ 9.3974
\citep{zeldovich1967,melosh1989}. Condensation of droplets occurs when
the cloud expands to the point where its pressure $P(T)$ is equal to
the saturation value $P_{\rm sat}(T)$. For typical parameters, this
crossover occurs at temperature $T=3000-4000$ K.} 
After the cloud reaches this threshold, the super-cooled vapor
will readily nucleate droplets, but their size (at a given time) is
limited by the Weber number, as outlined below.

\subsection{Nucleation of Droplets in an Expanding Vapor Cloud}
\label{sec:nucleate} 

The characteristic droplet size $s$ is given by the competition
between the surface tension $\surf$ of the liquid silicate and the
turbulent shearing that tends to destroy the droplet (where the
surface tension $\surf \approx 300$ dyne/cm; \citealt{walker1981}).
This balance is codified by the Weber number $\weber$,
which is given by 
\be
\weber = {\rho_{\rm g} v_{\rm g}^2 s \over\surf} \,,
\ee
where $\rho_{\rm g}$ is the gas density, $v_{\rm g}$ is the relative
speed between the droplet and the gas (basically the turbulent
velocity), and $s$ is the size of the droplet. In the limit
$\weber\gg1$, the flow stretches the fluid, and ram pressure forces
destroy forming droplets. In the opposite limit $\weber\ll1$, surface
tension dominates.  One expects $\weber\sim1$, as fragmentation of
larger droplets reduces their size down to this limit, and merging of
smaller droplets occurs up to this limit
\citep{hinze1955,villermaux2007,eggers2008}. As a result, the
characteristic droplet size at at given time takes the form
\be
s = {\surf \over \rho_{\rm g} v_{\rm g}^2} \,,
\label{size} 
\ee
where the density is given by equation (\ref{density}). The turbulent
velocity $v_{\rm g}$ is expected to comparable to the flow speed, so
that $v_{\rm g}\sim3$ km/s. Note that the size scale of equation
(\ref{size}) is often called the Hinze scale.

After combining the results given above, the size of the droplets
produced at a given epoch becomes 
\be
s = {\surf\over v_{\rm g}^2} {4\pi v^3 t^3 \over 3M}
\approx {4\pi \surf v t^3 \over 3M}\,,
\label{rocksize} 
\ee
where the final approximate equality assumes that $v_{\rm g}\sim v$
(the expansion speed is comparable to the turbulent speed).  Putting
in canonical numbers we find 
\be
s = 1.3 \,{\rm cm} \,
\left({t\over3\times10^6{\rm sec}}\right)^3\,.
\label{dropsize} 
\ee
The droplet size, and hence the rock size, thus depends sensitively on
the time (due to the rapidly decreasing gas density).  The initial
droplet size will be extremely small, but the {characteristic}
size grows as the gas cloud expands.

The droplets will have a distribution of sizes, determined by
the competition between merging, fragmentation, and other factors.
For example, simulations of giant impacts on the planet Mercury 
produce analogous silicate clouds from collision speeds in the range
$\sim20-30$ km/s \citep{benz2007}. The resulting size distribution has
a range $s=10^{-6}-20$ cm and a peak near $s\sim1$ cm (see also 
\citealt{benz1999}). Significantly, most of the mass resides in
bodies with sizes near the peak at $\sim1$ cm. 

{This scenario requires the droplets to remain (primarily) in 
liquid form, so they can collide and merge. We thus consider the
balance between evaporation and condensation.  After the saturation
threshold is reached, the cloud reaches a state where $P(T)\approx
P_{\rm sat}(T)$ as the cloud expands and the temperature falls
(\citealt{zeldovich1967,melosh1989}; see also \citealt{choksi2021}).
The cloud will be mix of vapor and droplets with a mass fraction
$\fdrop$. In this state, the droplets experience both evaporation and
growth by additional condensation. The net mass flux onto a droplet of
size $s$ can be written in the form
\be
{\dot m}_{\rm net} = 4\pi s^2 c_m \sqrt{ {Am_p\over2\pi kT}}
\left(P - P_{\rm sat}\right) = 4\pi s^2 c_m
\sqrt{ {Am_p\over2\pi kT}} \delta_p P_{\rm sat} \,,  
\ee
where $c_m$ is a dimensionless constant of order unity, $Am_p$ is the
mass of the silicate molecules, and the final equality defines the
fractional difference from equilibrium $\delta_p$. If the background
pressue is exactly equal to the saturation value, then the net mass
flux in and out of a droplet vanishes.  As the cloud expands, the
pressure decreases and the temperature falls. The saturation pressure
$P_{\rm sat}(T)$ is a steeper function of temperature, compared to
$P(T)$, so the cloud can enter a state where $P>P_{\rm sat}$. In
this case, the droplets experience a net gain in mass. The resulting
condensation releases latent heat, which increases the background
temperature, but lowers the vapor density, which decreases the
pressure. The net effect is to push the cloud {\it toward} the state
of equality $P\approx P_{\rm sat}$.  On the other hand, if the cloud
expands too rapidly (compared to equilibrium) so that $P<P_{\rm sat}$,
then mass flows from the droplets back into the background, thereby
producing net evaporation. This action leads to the consumption of
latent heat, which decreases the temperature of the background. But
since $P_{\rm sat}(T)$ is a steeper function then $P(T)$, the pressure
is moved back {\it toward} the saturation value.}

{The equilibrium is thus stable, as long as it can be maintained,
which requires the net evaporation time $t_{\rm net}$ to be shorter
than the expansion time. For example, for temperature $T=3000$ K, the
droplets have size $s\sim1\mu$m, so that the time scale becomes
$t_{\rm net}$ = $m/{\dot m}_{\rm net}\approx(8\times10^{-5}/\delta_p)$
sec. If the equilibrium conditions are met at the 1\% level, then
$t_{\rm net}\sim 0.008$ sec, which is much shorter than the expansion
time. When the temperature drops down to $T\approx1800$ K, where the
characteristic size $s\sim1$ cm, the droplets start to solidify
(and $t_{\rm net}$ is about 1/3 of the expansion time). As a result,
the droplets can be partially solid before they finish merging.} 

{In order for the cloud to maintain $P\sim P_{\rm sat}$ as the
temperature falls, the fraction of the mass in the vapor phase in the
background cloud must decrease, so that droplet fraction $\fdrop$
grows with time. The energy budget thus determines the final value of
the droplet fraction $\fdrop$. When integrated over the saturated path
given by $P(T)\approx P_{\rm ast}(T)$, the $PdV$ work for a parcel of
cloud material is comparable to the energy release from the latent
heat of vaporization.  As a result, the fraction $\fdrop$ is expected
to reach a few tenths by the time $s\sim1$ cm.}

\subsection{Opacity and Optical Depth}
\label{sec:opacity} 

The opacity of the gas --- and its optical depth --- affect the
long-term evolution of the expanding vapor cloud and can be estimated
as follows. Here we assume that opacity is dominated by the nucleating
droplets, as any pre-existing silicate dust is readily vaporized by the
collision. Suppose that a fraction $\fdrop$ of the mass has condensed
into droplets of size $s$ (where the size distribution is considered
below). The opacity of the expanding gas cloud takes the form 
\be
\kappa = \fdrop {3\over4\rho_\ell s} \,, 
\ee
where $\rho_\ell$ is the density of the silicate in liquid form.
The optical depth $\tauop$ of the expanding gaseous sphere is
then given by 
\be
\tauop = \kappa \rho_{\rm g} R = \fdrop {3\over4\rho_\ell s} \cdot
{3M\over4\pi R^2} = \fdrop {9\over16\pi} {M\over\rho_\ell sR^2}\,. 
\ee
Using the expression (\ref{rocksize}) for the droplet size $s$ derived
previously, the optical depth becomes  
\be
\tauop = {27\fdrop\over64\pi^2}
{M^2 v_{\rm g}^2 \over \rho_\ell \surf R^5}\,,
\ee
where $\surf$ is the surface tension of the liquid silicate. 
Here we are implicitly assuming that the droplets keep growing,
so that at given time $t$, when the cloud has size $R=vt$, the
droplets have grown to their maximum size at that epoch. The
opacity and the optical depth thus decrease rather rapidly. 
Putting in numbers, we find
\be
\tauop \approx \left({R \over 0.14\,{\rm AU}}\right)^{-5} \,,
\ee 
where we have taken the fraction $\fdrop=0.1$. In the absence of other
considerations, the cloud would thus remain optically thick until it
subtends a sizable fraction of its orbit, with a corresponding time scale
of about $7\times10^6$ sec ($\sim$ 0.22 year). In practice, however,
after the expanding gas cloud grows larger than its Hill radius, it no
longer follows the assumed spherical evolution (see below).

{Note that the expanding cloud remains optically thick long
enough (e.g., $\sim0.2$ yr) for the droplet radius to grow to $s\sim1$
cm (see equation [\ref{dropsize}]). As a result, the droplets can
potentially remain embedded in an optically thick, thermally coupled
cloud while they grow, and only later does the cloud become
transparent enough for them to cool and freeze.  This ordering is
important because the merging of droplets requires the silicate to
remain in liquid form. Merging would be compromised if the droplets
condense into solid rock, which could bounce or fragment instead 
\citep{benz1999,leinhardt2012}.} 

\subsection{Freezing Out the Growth of Droplets}
\label{sec:freezeout} 

In general, we expect that the droplets will merge and grow as the
gas cloud expands, because the lower density allows for ever larger
sizes.  This process cannot continue indefinitely, however, as the
merging droplets must be able to find each other to merge. The
interaction rate is set by the usual expresssion 
\be
\Gamma = n \sigma v_i \,,
\ee
where $n$ is the number density of droplets, $\sigma$ is their
effective cross section and $v_i$ is the effective speed of
interaction. Next we note that $n\sigma=\kappa\rho$, and the
latter quantity is already calculated. The interaction rate can
then be written in the form 
\be
\Gamma = \kappa\rho v_i = \kappa\rho R {v_i\over R}
= {\tauop \over t} {v_i\over v} \,. 
\ee
The expansion rate $H$, akin to a Hubble-like flow, is given by
\be
H \equiv {{\dot R} \over R} = {1 \over t} \,. 
\ee
The freeze out condition, $\Gamma=H$, thus occurs when the optical
depth of the cloud drops below unity (with the correction factor
$v_i/v$, which is expected to be less than unity). At this crossover
point, we can solve for the characteristic droplet size to obtain 
\be
s = \left({81\over256\pi}\right)^{1/5} \fdrop^{3/5}
M^{1/5} \surf^{2/5} v_{\rm g}^{-4/5} \rho_\ell^{-3/5}
\left({v_i\over v}\right)^{3/5} \,. 
\ee
Putting in numbers we find
\be
s_{\rm max} \approx 12\,{\rm cm} \,\,
\left({M \over 1 M_\oplus}\right)^{1/5}
\left({\fdrop \over 0.10}\right)^{3/5}
\left({v_{\rm g} \over 3\, {\rm km/s}}\right)^{-4/5}
\left({v_i\over v}\right)^{3/5} \,.
\label{smax} 
\ee
Note that the result depends on fractional powers of the
relevant input variables. In other words, the resulting maximum
droplet size is relatively insensitive to the total mass $M$
and the velocities $v$ and $v_i$.\footnote{Note that this derivation
assumes that the turbulent velocity $v_{\rm g}$ and the interaction
velocity $v_i$ do not vary with time.  One can also consider a model
where these speeds decay with time, where $v\sim1/t$ is a typical
result.  The resulting calculation produces similar results, but the
fractional powers would be different. } Keep in mind that equation
(\ref{smax}) represents the maximum size for droplets to grow; in
practice, the coagulation process will not always proceed to this
maximum state, as discussed in the following section.

\subsection{Outer Boundary}
\label{sec:boundary} 

The discussion thus far has allowed the expanding gas cloud to
evolve without interference from the background environment. In
practice, however, the Hill radius $R_H$ provides an effective
outer boundary to the expanding gas cloud, where 
\be
R_H = r \left({M \over 3M_\ast}\right)^{1/3} \sim 0.01 r \,, 
\ee
where $r\sim3-10$ AU is the orbital location. When the gas cloud size
$R \ll R_H$, it can expand without interference from the background
environment. In the opposite limit where $R\gg R_H$, tidal forces,
coriolis forces, and Keplerian shear act upon the expanding cloud and
lead to departures from the spherical model. An approximate boundary
is thus given by $R_H$.

Using the Hill radius as the outer boundary enforces a minimum gas 
density and a corresponding maximum droplet size  
\be
s_H = {4\pi\zeta r^3 \over 3M v_{\rm g}^2} {M \over 3M_\ast} =
{4\pi\zeta r^3 \over 9 v_{\rm g}^2 M_\ast} \approx 0.20 \,{\rm cm}\,\,
\left({r\over3\,{\rm AU}}\right)^3
\left({v_{\rm g}\over 3\,{\rm km/s}}\right)^{-2} \,,
\ee
where the final equality uses typical parameter values.  This
constraint is thus slightly stronger than the requirement that
the interactions between droplets freeze out. 

Next we note that the expanding cloud of droplets will continue
to expand after the freeze-out condition has been met. The droplets
will cool into rocky bodies with a characteristic (outward) radial
velocity $v\sim3$ km/s. Since the orbit speeds at the locations of
interest ($a$ = 1 -- 10 AU) are $v_{\rm orb}\sim15-30$ km/s, the rocks
will acquire orbits with moderate eccentricities. The rock velocities
are directed radially outward from the collision point, in the rest
frame of the cloud. In the center of mass frame, the rocky bodies have 
velocities corresponding to the orbit of the cloud center, with
additional velocity vectors of magnitude $v$ and random directions.
These velocity perturbations thus lead to a range of eccentricities.
The largest values are $e_{\rm max}\sim v/v_{\rm orb}$ = 0.1 -- 0.2.
The orbits will thus sample a range of radii centered on the star,
with $r \in [a(1-e),a(1+e)]$, or $(\Delta r)/a\sim0.1-0.2$.

\section{Poynting-Robertson Drag}
\label{sec:prdrag}

This section considers Poynting-Robertson (PR) drag
\citep{poynting1904} as a means of producing a debris disk around a
white dwarf during the final stages of stellar evolution, as an AGB
star transitions to become a white dwarf. We assume that the
collisions that produce the rocky bodies (Section \ref{sec:collide})
take place late enough that the intense mass loss phase from the AGB
wind is over. More specifically, we neglect the forces produced by
outflowing winds in this treatment, and consider only the effects of
radiation. This treatment includes time dependent stellar luminosity,
which is large at the time of the collisions and then decreases as
the star transitions into a white dwarf. 

\subsection{Orbital Evolution of Rocks from the PR Effect}
\label{sec:orbits} 

To start, we consider the PR effect including only the drag force
itself, and not the outward force due to radiation pressure. This
latter complication is considered in Appendix \ref{sec:orbradiation}.
This present approximation scheme is thus valid for sufficiently large
rocks. We also consider only circular orbits. As shown in Appendix
\ref{sec:orbeccentric}, nonzero eccentricity leads to correction terms
of (at most) $\sim3\%$, and the eccentricity decreases more rapidly
than the semimajor axis.

In this treatment, the magnitude of the force exerted on a rock of
{radius} $s$, with cross section $\sigma = \pi s^2$, is given by 
\be
F_{\rm PR} = {L_\ast \over 4\pi r^2 c} (\pi s^2) {v_{\rm orb}\over c} \,,
\ee
where the corresponding torque is given by
\be
T_{\rm PR} = r F_{\rm PR} = {L_\ast \over 4\pi c^2} (\pi s^2)
\Omega_{\rm orb}  = m {dJ\over dt} =
m \sqrt{GM_\ast r} { {\dot r}\over2r}\,.
\ee
Note that this formulation assumes that the rocks execute circular
orbits.  The mass of the rock can be written in the form $m$ =
$(4\pi/3)\rho s^3$, so that the equation of motion becomes 
\be
r {dr\over dt} = - {3L_\ast \over 8\pi\rho sc^2} \,.
\label{motionone} 
\ee

The luminosity of the star decreases with time. If we start
from the point of evolution when the star has maximum radius,
then the luminosity is given (approximately) by the expression
\be
L_\ast(t) = L_0 \left({t_0 \over t}\right) \,,
\label{lstart} 
\ee
where $L_0\approx10^4 L_\odot$ and $t_0\approx10^4$ yr. Note that this
approximation captures the leading order behavior of the dimming star
\citep{miller2016}, but this approach is readily generalized as shown
in Section \ref{sec:gentime}. Let us define dimensionless quantities 
\be
\xi = {r \over r_0} \,, \qquad \tau = {t \over t_0} \,, 
\qquad {\rm and} \qquad \swig = {s\over s_0} \,,
\label{defdimension} 
\ee
where $s_0=1$ cm. The equation of motion then takes the form
\be
\xi d\xi = - {\alpha\over\swig} {d\tau\over \tau} \,,
\qquad {\rm where} \qquad 
\alpha = {3L_0 t_0 \over 8\pi\rho s_0 c^2r_0^2} \,.
\label{eqmotion} 
\ee
If we take $\rho=2$ g cm$^{-3}$ and $r_0$ = 1 AU,
then $\alpha \approx 3.7$.

The equation of motion (\ref{eqmotion}) can be integrated to find
\be
\xi_0^2 - \xi_f^2 = \left({2 \alpha \over \swig}\right)
\ln(\tau/\tau_1) \,, 
\ee
where $\tau_1$ is the starting (dimensionless) time. 
The location of the rocks as they drift inward is then given by 
\be
\xi = \xi_0 \left[ 1 - \left({2\alpha\over\swig\xi_0^2}\right)
\ln(\tau/\tau_1) \right]^{1/2}\,. 
\ee
In order for the rocks to travel all the way to $\xi=0$, the following
requirement must be met
\be
\swig \xi_0^2 = 2\alpha\ln(\tau/\tau_1) \approx 68\,,
\ee
where the final equality assume fiducial values and a dimensionless
times $\tau/\tau_1=10^4$ ($t=10^9$ yr if the collision takes place
at $t=10^5$ yr). For this example, if we take $\swig=1$, we find that
rocks with size $s=1$ cm will travel inward from a starting location
at $\sim8$ AU in a time span of 1 Gyr. Smaller rocks will be moved more
efficiently, but only rocks large enough to avoid radiation pressure
will move inward (rather than outward). We thus find that if planets
collide just outside the photosphere of an AGB star, from 1 to 10 AU,
then PR drag can move them inward to the vicinity of the white dwarf
{(see also \citealt{wyatt2005,veras2015}).} 

Note that the radius of the star must decrease fast enough that the
inward moving rocks do not catch up to the shrinking stellar surface.
This requirement is readily satisfied, as shown by the following
consistency check: We need the rocks to move all the way to the center
of the system over an evolutionary time $t\sim1-10$ Gyr, so $\tau$
$\sim 10^5 - 10^6$. In addition, we expect the star to become a white
dwarf, with white dwarf dimensions, over a time $t\sim10^6$ yr or
$\tau\sim100$. If the location $\xi$ is outside the original radius of
the AGB star at this time, then $\xi>1$ at $\tau=100$ (where this
constraint is sufficient, but not necessary). For a starting time
$\tau_1$, these two constraints imply 
\be
1 + (2\alpha/\swig) \ln (10^2/\tau_1) \lta 
\xi_0^2 \lta (2\alpha/\swig) \ln (10^6/\tau_1) \,. 
\ee
For starting time $\tau_1=10$, for example, this constraint places the
initial location of the planet collision in the range $\sim4-10$ AU,
depending on late-time WD cooling, and where we have taken $\swig=1$.
The range of allowed locations moves outward for smaller rock sizes,
{and also varies with the collision time $\tau_1$}.

{Another related effect is that the inspiralling rocks generally
catch up with the shrinking silicate sublimation front, where they
evaporate into gaseous form. Such evaporation must occur in any model
involving rocky debris, as the stellar surfaces (of order $10^4$ K)
are above the sublimation temperatures ($\sim1500-2000$ K). As shown
previously, the incoming rocks reach the sublimation temperatures at
radial scales of order $\sim1R_\odot$ (e.g., see Section 2.5 of
\citealt{veras2022}).  The timing and location of the evaporation
depends on many factors, including the timing and location of the
original collisions, the rock size, and their composition. Another
complication is that the gaseous material can self-shield, so that 
the evaporation front moves inward \citep{rafikov2012}. In any case,
the pollution of the white dwarf ultimately takes place while the
silicate material is in vapor form \citep{rafikov2011,veras2022}. In
this work, we assume that rocks delivered to the sublimation region
can eventually accrete onto the white dwarf (although future work
should consider this process in greater detail). }

\subsection{Constraints}
\label{sec:constraint}

We first consider the collision rate for rocks in the disk of material
just outside the star. The rate $\Gamma$ is given by
\be
\Gamma = n \sigma v_{\rm g} = {\Sigma \over m H} \sigma v_{\rm g} \,,
\ee
where $\Sigma$ is the surface density, $\sigma$ is the cross section,
and the scale height $H = v_{\rm g}/\Omega$. Here the speed $v_{\rm g}$
is the random (turbulent) velocity of the rocks. Using the geometric
results $\sigma=\pi s^2$ and $m=(4\pi/3)\rho_s s^3$, the collision
rate becomes  
\be
\Gamma = {3\Sigma\Omega \over 4\pi\rho_s s} 
\ee
The migration rate from PR drag can be written in the form
\be
\gamma = {3L_\ast \over 8\pi \rho_s s c^2 r^2} \,. 
\ee
We can define a benchmark surface density by setting the migration
rate equal to the collision rate {(see also \citealt{wyatt2005})}.
This procedure implies 
\be
\Sigma_X = {L_\ast \over 2\Omega c^2 r^2} \sim
0.0005 \,{\rm g}\,\,{\rm cm}^{-2} \,,
\label{sigmax} 
\ee
where the final approximate equality uses canonical parameter values
($L_\ast$ = $10^4L_\odot$, $r=1$ AU, $M_\ast=1M_\odot$). If the
surface density is larger than this threshold, then rocks can collide
before they move. This benchmark surface density $\Sigma_X$ implies a
corresponding scale $M_X$ for the maximum mass of material that can be
delivered to the white dwarf without collisions.  Since the starting
distances are not expected to be greater than $\sim10$ AU, this mass
threshold is given by 
\be
M_X = \pi (10 {\rm AU})^2 \Sigma_X \approx 3.3 \times 10^{25}
\, {\rm g} \, \approx 0.006 M_\oplus\,, 
\ee
which is somewhat larger than that of the asteroid belt
\citep{krasinsky2002}. {This mass scale falls within
the mass range inferred from observations of WD pollution, 
$3\times10^{22}-3\times10^{26}$ g, but is smaller
than the largest value.} 

In general, we expect the initial mass of the debris disk to be larger
than $M_X$, so that collisions will take place. The largest bodies
will collide and shatter, whereas the smallest bodies will collide and
experience sticking, bouncing, erosion, abrasion, and cratering
\citep{blum2008,birnstiel2024}. {In addition, radiation pressure
will remove the smallest rocks, where the size threshold depends on
the time of the collisions. Another issue is that the rocks emerging
from the initial cloud will not adjust immediately to a disk
configuration \citep{jackson2012}.  The orbiting rocks tend to return
to the collision point and thus experience larger densities than the
above estimates. On the other hand,} inelastic damping from the
collisions will rapidly dissipate the random velocities of the rocks
and thereby produce a thin, cold, shear-dominated disk, thereby
reducing the collision rate, {analogous to the case of the rings
of Saturn \citep{cuzzi2009}.\footnote{In addition, debris disks
associated with main-sequence stars show evidence that some material
can survive in highly collisional disks \citep{sommer2025,wyatt2025}.}
For example, experiments show that only about 1\% of the collision
energy is transferred to the kinetic energy of the largest fragments,
which carry the majority of the mass \citep{michikami2007} The rocks
thus execute a cascade, starting with sizes $s\sim1$ cm and speeds
$v\sim1$ km/s, which become $s\sim$ few mm and $v\sim50$ m/s after one
round of collisions (see also \citealt{fujiwara1977,wickham2018}).
Subsequent collisions reduce the speeds further, but rocks with small
sizes and low speeds will no longer shatter, but can coagulate
instead.} These processes lead to a distribution of rock properties
with characteristic size of order $s\sim1$ cm.

{For reference, we can derive a benchmark mass accretion rate for
the limit of maximally efficient collisions. In this limit, we assume
that the surface density interior to the collision radius
$r_{\rm col}$ self-regulates toward $\Sigma_X$ (equation
[\ref{sigmax}]). In this limit, collisions act efficiently to shatter
the rocks until they reach blowout sizes and are expelled, while any
deficit is replenished by inward drift from the reservoir
\citep{wyatt2005}. In this saturated state, the throughput is
bottlenecked at the reservoir radius, with the characteristic mass
accretion rate
\be
\dot{M}_{\rm PR} \;=\; 2\pi r_{\rm col}\,\Sigma_X\,v_{\rm PR}
\;=\;\frac{\pi\,G M_\ast L_\ast(t)}{\Omega\,c^{3}\,r_{\rm col}^{2}}
\;=\;\frac{\pi\sqrt{G M_\ast}\;L_\ast(t)}{c^{3}\sqrt{r_{\rm col}}}.
\label{mdotgrind} 
\ee
In this approximation, the drift speeds correspond to blowout
sizes ($\beta=1/2$; see Appendix \ref{sec:orbradiation}) so that 
$v_{\rm PR}$ = $GM_\ast/(c\,r_{\rm col})$ and $\Omega$ is evaluated at
$r_{\rm col}$. Note that this result is independent of disk mass,
particle size, and and material density. The benchmark mass accretion
rate $\dot{M}_{\rm PR}$ is set by stellar parameters (and
$r_{\rm col}$), and tracks the stellar luminosity $L_\ast(t)$. In this
limit, the decreasing $L_\ast$ is partially offset by decreasing rock
size. This approximation breaks after the minimum rock size (set by
$\beta=1/2$) reaches the limit of interstellar dust grains
($s\sim0.1\mu$m).}

Next we consider the optical depth of the rocky material, which
is given by 
\be
\tauop = n \sigma \ell = {3\Sigma \ell \over 4\rho_s s H} \, .
\ee
If we require that $\tauop<1$, so that stellar photons can reach
the rocks to drive the PR effect, then we obtain the constraint
\be
3 \Sigma \ell < 4\rho_s s H \, .
\ee
This constraint can be written
\be
\Sigma \lta (3\,{\rm g}\,{\rm cm}^{-2})\,
\swig {H\over\ell} \qquad {\rm or} \qquad
M_{\rm max} \lta 10 M_\oplus \,, 
\ee
where we have used standard values. The quantity $H/\ell$ is the
essentially the ratio $H/r$ for the rocky disk (the path length $\ell$
should be comparable to the radial location $r$), where the mass
estimate uses $H/\ell=0.1$ (see Section \ref{sec:boundary}). In any
case, for the expected masses, the debris disk is expected to remain
optically thin so that its constituent rocks are exposed to stellar
photons.

{Another constraint arises from the requirement that the rocks
produced from the planetary collisions do not immediately evaporate
when exposed to the stellar radiation field.  The sublimation radius
for silicates is given by
\be
r_{\rm sub}\approx3.6 {\rm AU}\,\,
\left({L_\ast\over 10^4L_\odot}\right)^{1/2}
\left({T_{\rm sub}\over1500{\rm K}}\right)^{-2}\,.
\ee
With the maximum stellar luminosity, this constraint requires the
collisions to take place outside $2-4$ AU, depending on the
sublimation temperature, which depends on the type of silicate.  As
shown below, the preferred collision times correspond to $\tau\sim10$
when the stellar luminosity is smaller, so that the sublimination
radius $r_{\rm sub}\lta1$ AU, whereas the collision radii
$r_{\rm col}=\xi_0r_0\gta6$ AU.} 

\section{Surface Density Evolution}
\label{sec:evolve}

With the motion of individual rocks specified, the next step is to
consider the evolution of the entire field of rocky bodies, as
described by its surface density. We first consider the surface
density evolution for rocks bodies of a single size, and then
generalize the results for a distribution of sizes. 

\subsection{Surface Density for a Single Rock Size}
\label{sec:sigmaone} 

The continuity equation for the surface density has the form
\be
{\partial\Sigma\over\partial t} + {1\over r}
{\partial\over\partial r} \left(ru\Sigma\right) = 0\,,
\ee
where $u$ is the radial velocity. Here we use the results from the
previous section to mandate that the radial velocity is driven by PR
drag and thus has the form 
\be
u = {dr\over dt} = - {3L_\ast \over 8\pi\rho{s}c^2r} \,. 
\ee
Using the dimensionless variables variables defined through
equation (\ref{defdimension}), the continuity equation becomes
\be
\tau {\partial \Sigma\over \partial \tau} -
{\alpha\over\swig} {1\over \xi} 
{\partial\Sigma\over\partial\xi} = 0 \,. 
\ee
Since the speed $u$ depends on the size, the surface density is that
for rocks of size $\swig$ only, i.e., one obtains a continuity
equation of this form for each value of $\swig$.  Next we define a
logarithmic time variable 
\be
\eta = \alpha \ln (\tau/\tau_1) \,,
\ee
so that the evolution equation becomes
\be
{\partial\Sigma\over\partial\eta} = {1\over \swig\xi} 
{\partial\Sigma\over\partial\xi} \,.
\label{etaevolve} 
\ee
Along characteristics we have
\be
{d\xi\over d\eta} = - {1\over\swig\xi} \qquad \Rightarrow \qquad
X = \xi^2 + 2\eta/\swig = constant  \,. 
\ee
The solutions are thus of the form
\be
\Sigma (\xi,\eta) = F(X) \,,
\ee
where the form of $F$ is chosen to satisfy the initial conditions.
At our starting time,
\be
t=t_0, \qquad \tau=1, \qquad \eta=\alpha\ln(\tau/\tau_1)=0,
\qquad X = \xi^2\,. 
\ee
We can thus choose $F(X)$ to have the form
\be
F(X) = \exp[ - (\sqrt{X} - \xi_0)^2/\width^2 ] \,,
\ee
or equivalently 
\be
\Sigma_1 (\xi,\eta) =
\exp[ - (\sqrt{\xi^2+2\eta/\swig} - \xi_0)^2/\width^2 ] \,,
\label{solution} 
\ee
where $\Sigma_1$ defines the functional form for the surface
density of a single rock size. The full surface density can
be constructed from these building blocks, as outlined below. 

\subsection{Surface Density for Distribution of Rock Sizes}
\label{sec:sigmagen} 

This discussion thus far has assumed that the rocks have a single size
given by $\swig$, and then finds the surface density function denoted
here as $\Sigma_1$.  In actuality, the rocks have a distribution of
sizes. We can introduce a distribution by mass, 
\be
{dm\over d\swig} \qquad {\rm s.t.} \qquad
\int_0^\infty {dm\over d\swig} d\swig = 1 \,, 
\ee
so that the generalized expression for the surface density
is given by
\be
\Sigma(\xi,\eta) = \Sigma_0 \int_0^\infty {dm\over d\swig} d\swig 
\exp[ - (\sqrt{\xi^2+2\eta/\swig} - \xi_0)^2/\width^2 ] \,, 
\ee
where $\Sigma_0$ is a constant that is determined by the
total initial mass in rocks in the annulus of interest.

Note that this form for the surface density $\Sigma$ does {\it not}
obey the usual continuity equation for a two dimensional surface
density of a disk. The usual continuity equation assumes that the
fluid has a single velocity. In this case, however, the individual
contributions from each different rock size have a different inward
speed. The surface density contribution for a given rock size does
obey the continuity equation, by construction, but the full solution
does not.

The standard choice for the distribution of rock sizes by number
can be written in the power-law form
\be
{dn\over ds} \propto s^{-p} \,, 
\ee
where the distribution is determined by the condensation process of
Section \ref{sec:collide}. For comparison, the index $p\sim3.5$ for
interstellar dust grains, but is generally smaller in the
circumstellar environment (see the review of \citealt{birnstiel2024}).
For the sake of definiteness, this work uses $p=3$ {(consistent
with observations of Saturn's rings; \citealt{cuzzi2009})}, which implies
$dm/d\swig$ = $constant$. Here we consider the rocks to have maximum
size given by the fiducial size $s_0$ = 1 cm, so that the maximum
value of $\swig=1$. The rocks must also have a minimum size, and we
denote the minimum value of $\swig$ as $\smin$. Rocks smaller than the
size $\smin$ can be expelled by radiation pressure, where the value
depends on the time of the collision and hence the stellar luminosity
at that epoch {(again see Appendix \ref{sec:orbradiation})}. 
For the sake of definiteness, we take $\smin=0.1$, corresponding to a
rock size of 1mm, as expected for a starting time $\tau=10$,
{as the fiducial case. Other choices of $\smin$ can be readily
considered since we have mostly analytic solutions. Notice that at 
later times, smaller rocks can survive as the stellar luminosity
decreases, and can be produced via collisions of larger rocks. As long
as the mass is dominated by the larger bodies with $s\sim1$ cm,
however, the results are not overly sensitive to the choice of
$\smin$. }

{With the form of the size distribution specified, it can be} 
normalized according to 
\be
\int_0^\infty {dm \over d\swig} d\swig =
\int_{\smin}^1 {d\swig \over 1 - \smin} = 1\,. 
\ee
For this choice, we can carry out the integration to
find the total mass at the start of the disk's evolution, 
\be
M_0 = 2\pi r_0^2 \Sigma_0 \int_0^\infty \xi d\xi
\int_{\smin}^1 {d\swig\over 1 - \smin} 
\exp[ - (\xi - \xi_0)^2/\width^2 ] \,, 
\ee
which evaluates to the form
\be
M_0 = 2\pi r_0^2 \Sigma_0 \xi_0 \width \sqrt{\pi} 
+ {\cal O}\left({\rm e}^{-\xi_0^2/\width^2} \right) \,. 
\ee
Since we expect $\width/\xi_0\sim0.1$, the error term
$\sim {\rm e}^{-100} \ll 1$.

\subsection{Generalized Time Evolution}
\label{sec:gentime}

The discussion thus far considers the time dependence of the stellar
luminosity to have the form $L_\ast(t)\propto1/\tau$. However, we can
consider a general form for the time evolution where 
\be
L_\ast(t) = L_0 \fstar(\tau) \,. 
\ee
The time behavior $L_\ast\sim1/t$ is a good model for times
$t=t_0\to100t_0$ (recall that $t_0=10^4$ yr). After $\sim10^6$ years,
however, the star becomes a bonafide white dwarf and obeys a cooling
law with somewhat steeper time dependence. The power-law relation 
\be
L_\ast \propto \tau^{-7/5} 
\label{wdcool} 
\ee
provides a good starting approximation for $\tau>100$
\citep{mestel1952}. However, a number of more recent improvements to
white dwarf models lead to more complicated behavior. These updates
include neutrino cooling, which makes the initial phases suffer more
energy loss, and crystallization, which slows down the cooling rate at
ages of order $\sim1$ Gyr
\citep{fontaine2001,hansen2003,tremblay2019}.

For any choice of $\fstar(\tau)$, we can re-define the time coordinate
$\eta$ so that it takes the form
\be
\eta = \alpha \int^\tau \fstar(\tau') d\tau' \,. 
\ee
The differential equation (\ref{etaevolve}) that describes the
evolution of the surface density remains the same, and the form of the
solution (from equation [\ref{solution}]) also remains the same. The
only thing that changes is the interpretation of the meaning of the
time coordinate $\eta$.

Although the form of the solution is unchanged, this change of time
coordinate implies that the parameter $\eta$ generally has a finite
range. For the choice $\fstar=1/\tau$, where $\eta$ =
$\alpha\ln(\tau/\tau_1)$, $\eta$ can become arbitrarily large as
$\tau\to\infty$. In contrast, if we use the Mestel cooling law
(\ref{wdcool}) for $\tau>100$, then the mapping $\eta(\tau)$
takes the form  
\be
\eta/\alpha = \ln (100/\tau_1) + {5\over2} \left[ 1 -
\left({100\over\tau}\right)^{2/5}\right] \,. 
\ee
In the long time limit $\tau\to\infty$, the parameter $\eta$
{is} thus finite.  

This result -- a finite value of $\eta$ -- arises due to the total
(finite) amount of energy available to drive PR migration. For a given
time evolution function $\fstar(\tau)$, the total energy emitted by
the star is given by the integral 
\be
E_T = \int_{t_0}^t L_\ast(t) dt =
L_0 t_0 \int_1^\tau \fstar(\tau)d\tau = L_0 t_0 \eta/\alpha \,. 
\ee
The energy scale given by $L_0t_0$ = $4\pi\times10^{48}$ erg.
The total bound state energy of a cold white dwarf, considered
here to be a perfect $n=3/2$ polytrope, has the form
\be
E_{\rm WD} = - {3\over7} {GM_\ast^2\over R_\ast} \approx
2 \times 10^{50} \, {\rm erg}\,. 
\ee
This energy scale is approximately the total energy store that is
available to be radiated away during the entire history of cooling,
from the AGB state to that of the cold white dwarf. We thus obtain
\be
E_T = L_0 t_0\eta/\alpha\approx E_{\rm WD} \approx 10 L_0 t_0
\qquad \Rightarrow \qquad \eta_{\rm max} \approx 10\alpha
\lta40\,.
\label{etamax} 
\ee
The value of $\eta$ is thus bounded from above via equation
(\ref{etamax}), with a working upper limit of
$\eta_{\rm max}\approx40$.  In practice, however, the largest value of
$\eta$ that can be realized will be smaller --- the collisions can
take place after $t_0$ and some energy is lost before the above
accounting starts. As a result, most evolutionary scenarios will have
a smaller range with $\eta_{\rm max}\sim20-30$.

Since the total distance $r_0$ that rocks can be transported is
constrained by the amount of available radiation, so that
$r_0^2 \propto \int L_\ast dt$ (Section \ref{sec:orbits}), this
upper limit on the energy implies a corresponding limit on rock
transport (see also Appendix \ref{sec:upperlimit}).

\subsection{Results}
\label{sec:results}

This section presents the time evolution for the surface density, the
debris disk mass, and the mass accretion rate as a function of the
time parameter $\eta$. Although the solutions for $\Sigma(\eta)$,
$M(\eta)$, and $dM/d\eta$ apply for any luminosity history, the
mapping between the temporal parameter $\eta$ and the physical time
$t$ depends on the white dwarf cooling law, as outlined in the
previous section. We show the mass accretion rate ${\dot M}$ for a
representive case. For the sake of definiteness, the range of rock
sizes is taken to be $\smin=0.1 \le \swig \le 1$.

\begin{figure}
\centering
\includegraphics[width=0.80\linewidth]{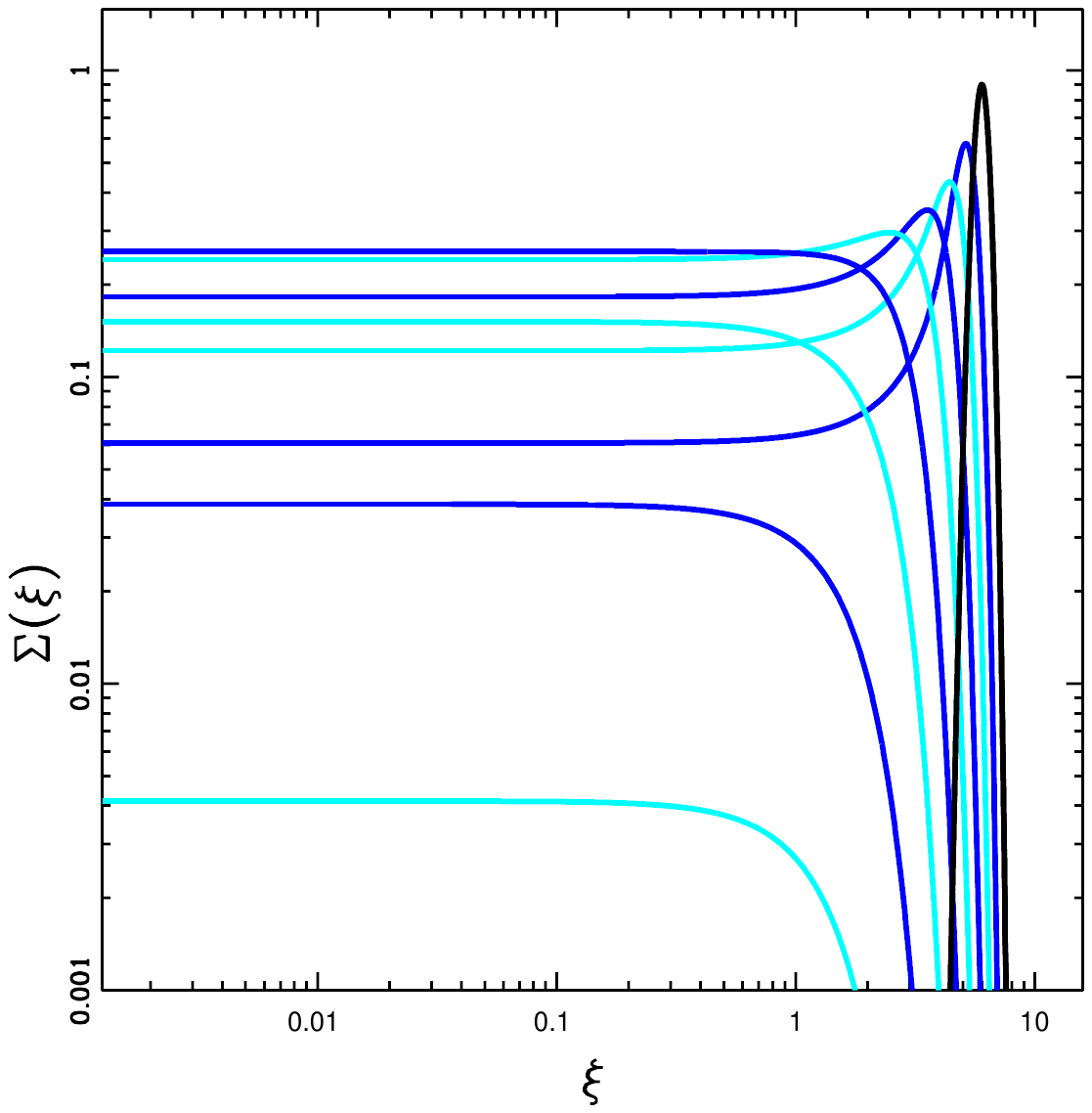}
\vskip-1.0truein
\caption{Surface density distributions of rocky material for a
collection of time parameters from $\eta=0$ (right) to $\eta=24$
(bottom). The collision is assumed to take place at dimensionless
radius $\xi_0=6$, where the initial rock distribution has width
$\width=\xi_0/10$, as shown by the black curve. The peak of the
surface density moves inward with time. The dark and light blue
curves show the subsequent evolution in equal increments of $\eta$. }
\label{fig:sigma}
\end{figure} 

The evolution of the surface density is depicted in Figure
\ref{fig:sigma}. For this example, the collision takes place at
dimensionless radius $\xi_0=6$ and evolves with time as shown. The
initial distribution is confined to an annulus near the collision
site, and the peak of the distribution moves inward with time.  Note
that the surface density in the inner region first increases with time
(since the initial distribution is confined to an annulus in the
vicinity of the collision), and then later decreases with time (as the
rock supply in the disk becomes depleted). Surface density
distributions are shown for values of $\eta$ = 0 -- 24, in equal
increments. This lowest curve corresponds to $\eta=24$, which is close
to the maximum expected value (Section \ref{sec:gentime}). Since
we expect the collision to take place at $\tau_1\sim10$, just after
the AGB winds have finished removing mass from the star, final value
of $\eta$ could be somewhat lower ($\eta_{\rm max}\approx18$ for
$\tau_1$ = 10 and pure Mestel cooling).  In any case, the surface
density will asymptotically approach one of the later curves shown
here.  Notice also that the evolution proceeds more slowly for larger
values of the starting radius $\xi_0$.

\begin{figure}
\centering
\includegraphics[width=0.80\linewidth]{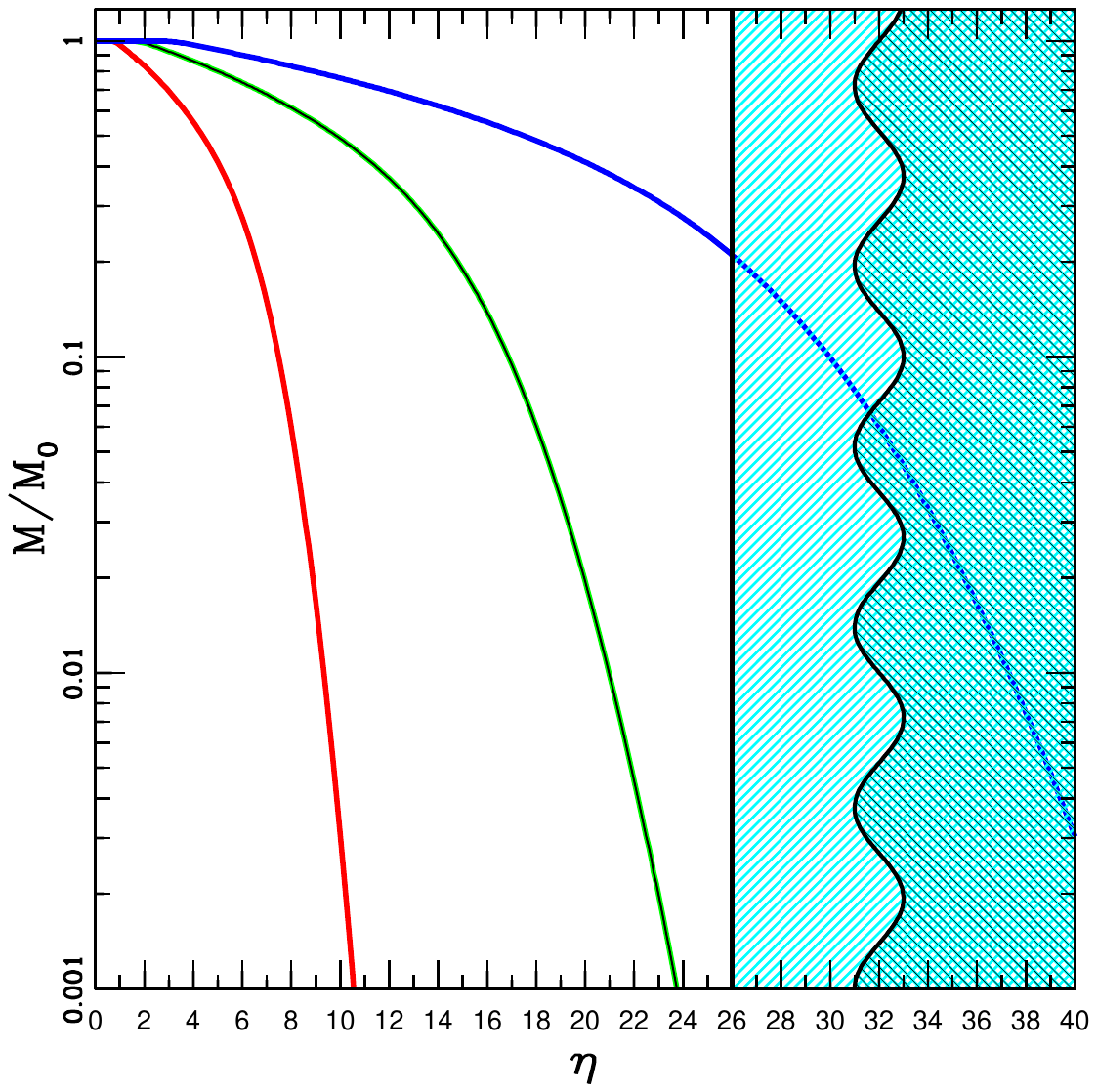}
\vskip-1.0truein
\caption{Mass remaining in the debris disk as a function of the
temporal parameter $\eta$. The collision is assumed to take place at
dimensionless radius $\xi_0$, where the initial rock distribution has
width $\width=\xi_0/10$. Results are shown for starting locations
$\xi_0$ = 4 (red), 6 (green), and 8 (blue). The mapping of the time
parameter $\eta$ to physical time depends on the details of white
dwarf cooling (Section \ref{sec:gentime}). In general, $\eta$ has a
finite range, so that large values of $\eta$ are not reached, as
indicated by the shading on the right side of the figure. The limit
shown at $\eta=26$ corresponds to Mestel cooling over the full allowed
range of physical time ($1\le\tau\le\infty$). If collisions occur
later ($\tau_1>1$), the range in $\eta$ is correspondingly smaller.
The darker shading (right) depicts the limit for more physical (but
uncertain) white dwarf cooling evolution. }
\label{fig:masstime}
\end{figure}

The mass of the debris disk is shown as a function of time in Figure
\ref{fig:masstime}. The fraction of the disk remaining at a given
epoch depends primarily on the location of the collision, which sets
the starting location for the annulus of rocky material. The figure
shows the time evolution of the mass fraction for starting radii
$\xi_0$ = 4 (red), 6 (green), and 8 (blue). As shown in Appendix
\ref{sec:halflife}, the shape of the curves depends on the composite
variable $h=2\eta/\xi_0^2$, and the time required for half of the disk
mass to be accreted is given by $\eta_{1/2}\approx0.274\xi_0^2$.  The
debris disks thus retain a substantial fraction of their mass for
billions of years, provided that the collisions take place at
$r\sim5-10$ AU. The time parameter $\eta$ generally has a finite range.
Although the limiting value depends on the details of white dwarf
cooling, the shaded region (right side of the figure) depicts the
maximum value for pure Mestel cooling (vertical line) and the upper
limit given by energy considerations (wavy curve). 

\begin{figure}
\centering
\includegraphics[width=0.80\linewidth]{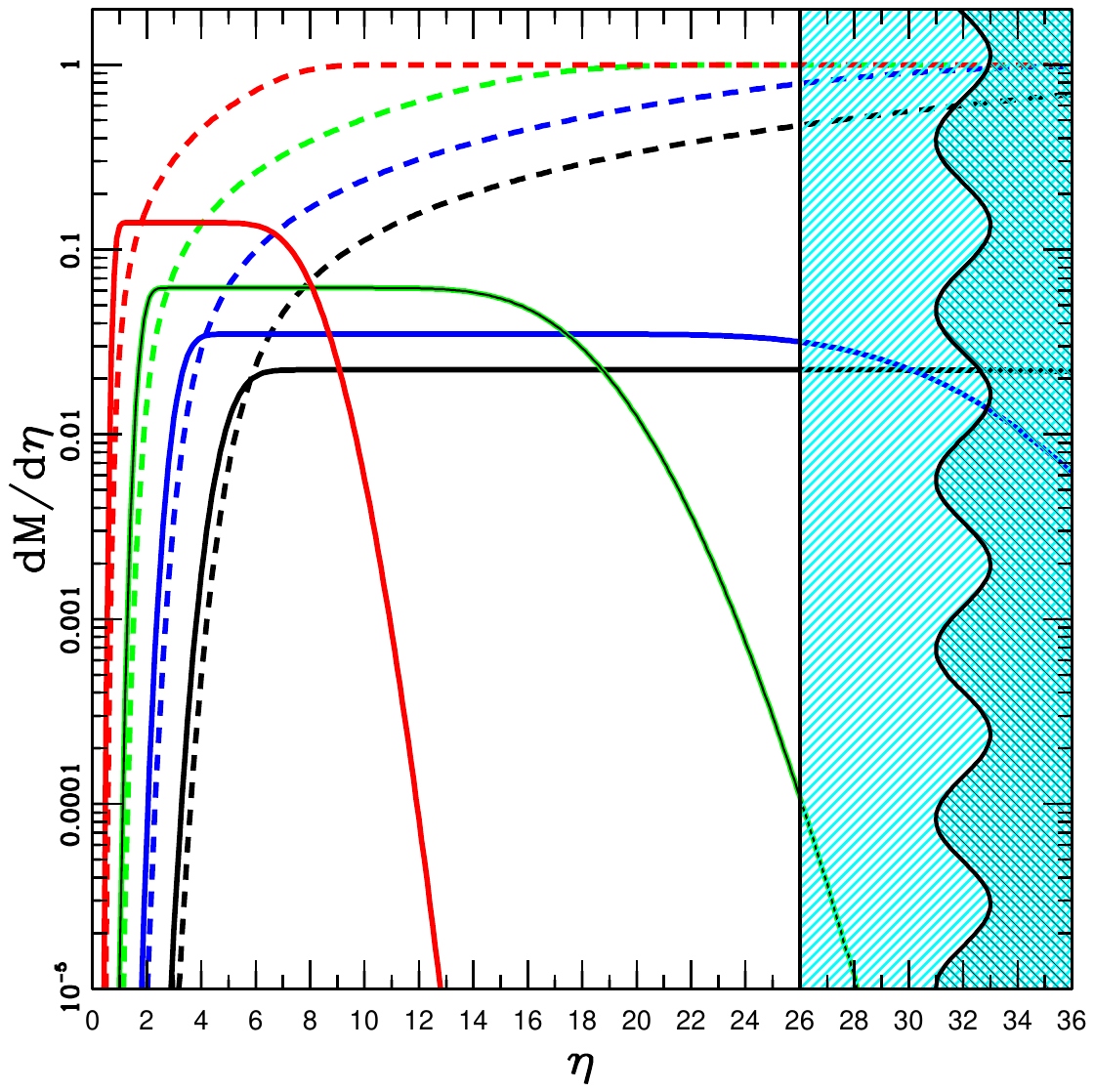}
\vskip-1.0truein
\caption{Reduced mass accretion rate $dM/d\eta$ as a function of the
temporal parameter $\eta$ (shown by the solid curves). The rate
$dM/d\eta$ shown here is normalized by the initial disk mass $M_0$.
The second set of increasing dashed curves shows the cumulative mass
that is accreted as a function of $\eta$, where all of the mass is
accreted in the limit $\eta\to\infty$. The collision is assumed to
take place at dimensionless radius $\xi_0=4$ (red), 6 (green), 8
(blue), and 10 (black). In general, $\eta$ has a finite range, as
depicted by the shading on the right side of the figure (see text and
the caption to Figure \ref{fig:masstime}). The physical mass accretion
rate $dM/dt$ can be obtained by multiplying by the factor $d\eta/dt$,
which depends on the luminosity history of the star (e.g., see
Figure \ref{fig:mdot}). }
\label{fig:meta}
\end{figure}

In this scenario, the surface density distributions decrease with
time, but remain flat at small radii (see Figure \ref{fig:sigma}).
However, current observations detect debris disks around only a small
fraction of white dwarfs, and infer a limited radial extent
$r_d\lta1R_\odot$ (e.g., \citealt{farihi2016}). One way to understand
this finding is that the disks become optically thin (the flux density
$F_\nu\sim\tauop$). For example, for the maximum inferred disk mass
($M=3\times10^{26}$ g), disk radius $r_d=1$ AU, and rock size $s=1$
cm, the optical depth $\tauop\approx0.16<1$. The disk mass decreases
with time (Figure \ref{fig:masstime}), so that $\tauop\ll1$, making
the disks more difficult to detect. {However, the limited observed
extent of white dwarf disks could arise from an actual compact
physical configuration, and this truncated geometry poses a challenge
for this scenario. Note that this same issue arises for the competing
model of eccentric dynamical delivery, as the required debris streams
are also extended beyond $\sim1R_\odot$.  In any case, the level of
infrared excesses in these systems provides important constraints on
the theoretical models. }

\begin{figure}
\centering
\includegraphics[width=0.80\linewidth]{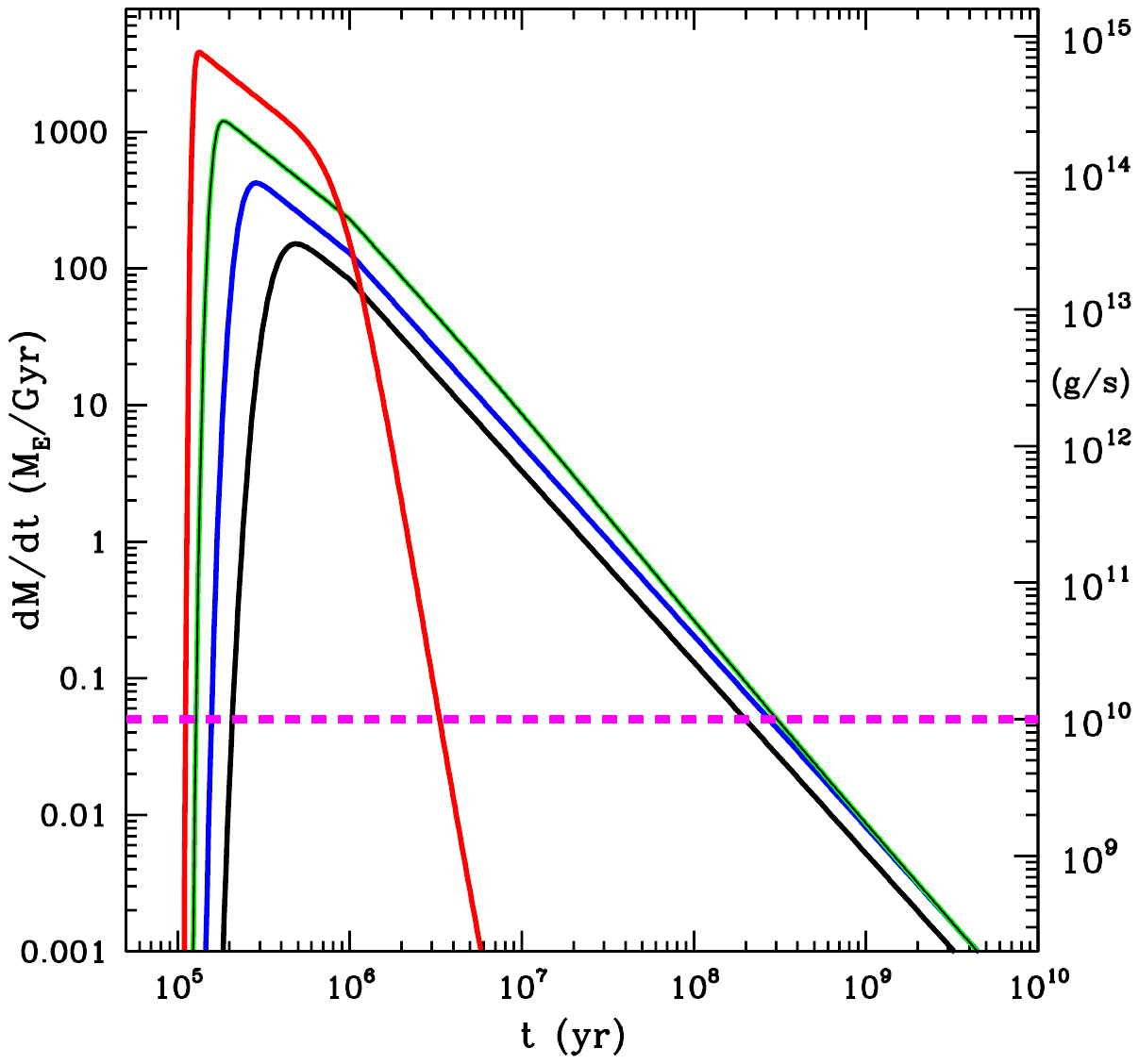}
\vskip-1.0truein
\caption{Mass accretion rate $\dot M$ as a function of time (in
years). The accretion rate is given in units of $M_\oplus$/Gyr
{on the left vertical axis and in units of g/s on the right
axis. (Note that 1 $M_\oplus$/Gyr = $1.9\times10^{11}$ g/s.)} 
The curves show results for different collision locations, as
specified by the dimensionless radius $\xi_0=4$ (red), 6 (green),
8 (blue), and 10 (black). The collision is assumed to occur at
time $t=10^5$ yr with starting mass $M_0=1M_\oplus$. {The
horizontal dashed line depicts the maximum observed mass
accretion rate.} } 
\label{fig:mdot}
\end{figure} 

The mass accretion rate onto the central white dwarf can be
written in the form
\be
{\dot M} = 2\pi r_0^2 \Sigma_0 \left({d\eta\over dt}\right) 
{1\over1-\smin} \int_{\smin}^1 {d\swig\over\swig} 
\exp[ - (\sqrt{2\eta/\swig} - \xi_0)^2/\width^2 ] =
{dM\over d\eta} {1\over t_0}\left({d\eta\over d\tau}\right)\,, 
\ee
where the sign is chosen so that mass accretion onto the star is
positive. The quantity $dM/d\eta$ can be determined for any cooling
law (equivalently $L_\ast(t)$ or $f_\ast(\tau)$), but finding the mass
accretion rate $dM/dt$ itself requires specification of $\eta(t)$.
Figure \ref{fig:meta} shows how the reduced mass accretion rate
$dM/d\eta$ varies with the temporal parameter $\eta$. The accretion
rate is given in units of $M_0$, where $M_0$ is the initial disk mass.
The curves show $dM/d\eta$ for varying collision locations, which set
the starting point for the debris field (here $\xi_0$ = 4, 6, 8, and
10).  At early times, rocks have not yet reached the central region,
so that the accretion rate is initially zero. The accretion rate then
rapidly approaches a nearly constant level.  This behavior continues
until the peak of the surface density distribution reaches the
origin. After that epoch, the accretion rates decrease rapidly with
time.

In order to find the mass accretion rate in physical units, we need to
specify the luminosity evolution of the star, which determines the
mapping $\eta(t)$. As a working example, we adopt the Mestel cooling
law of equation (\ref{wdcool}) to determine the factor
$d\eta/d\tau$. For this case, the starting mass of the debris disk
$M_0 = 1 M_\oplus$ and the time of the collision is taken to be
$\tau_1=10$ or $t_1=10^5$ yr.  The resulting mass accretion rate is
shown in Figure \ref{fig:mdot} for different collision locations
$\xi_0$ = 4 (red), 6 (green), 8 (blue), and 10 (black). The accretion
rate onto the star is zero until the smallest rocks can reach the
center. After that time, the accretion rate quickly increases to its
maximum and then steadily decreases. For the case $\xi_0=4$, most of
the disk mass is accreted before $t\sim10^6$ yr, while the stellar
luminosity remains high.  Since the stellar luminosity steadily
decreases, for cases with larger starting radii, the debris disk
lasts much longer. For ages of $t$ = 1 -- 5 Gyr, the mass accretion
rate $\sim0.001-0.01$ $M_\oplus$/Gyr $\approx 2\times10^8 -
2\times10^9$ g/s, consistent with observational estimates 
\citep{farihi2016}. {The mass accretion rates decrease with time
according to ${\dot M}\sim1/t$. This result is consistent with
observations, which find that mass accretion rates in WD systems
decrease as power-laws \citep{chen2019}, and recent constraints
indicate a decrease by a factor of $\sim10$ for ages 1 -- 8 Gyr
\citep{blouin2022}.}

\begin{figure}
\centering
\includegraphics[width=0.80\linewidth]{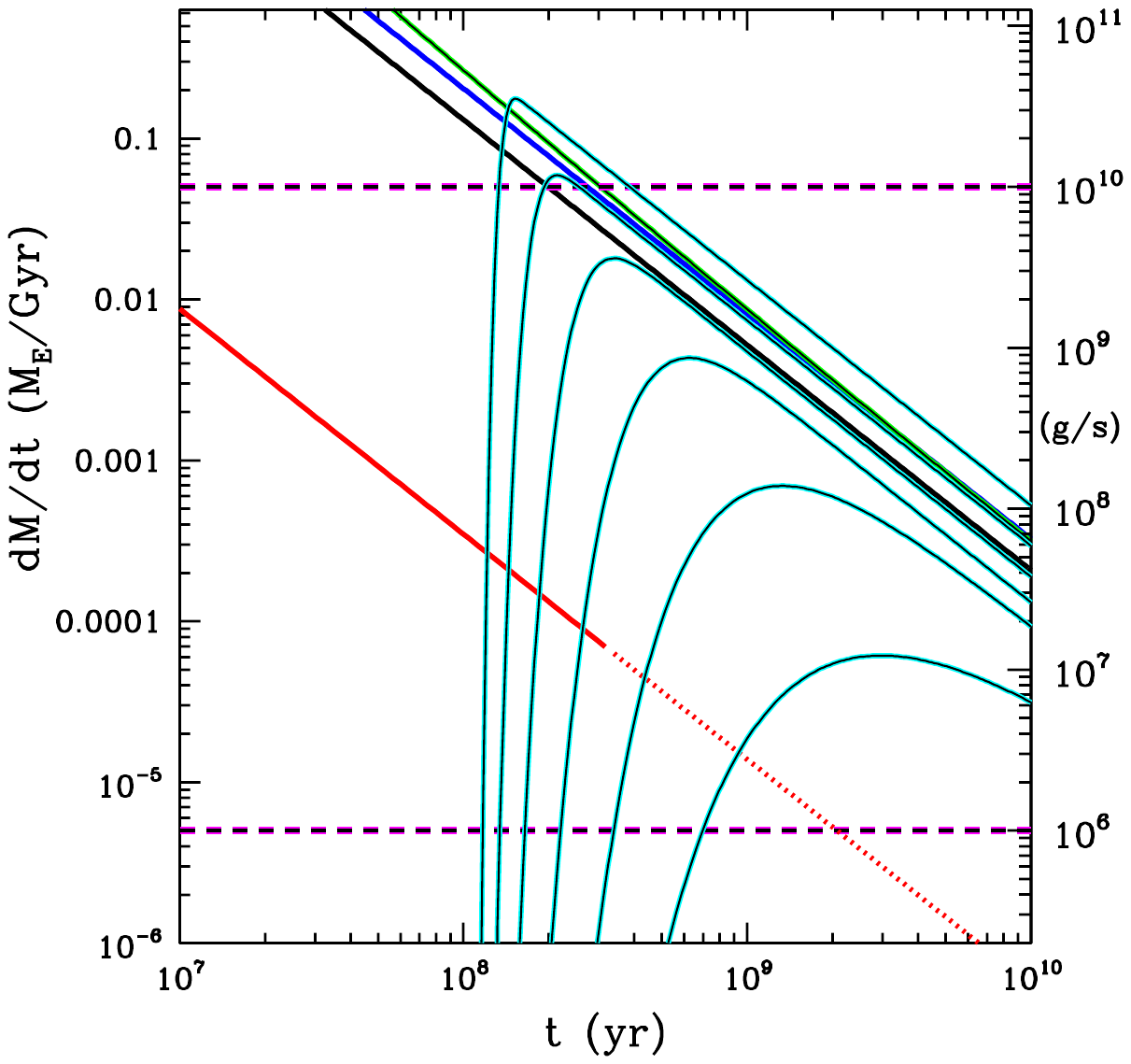}
\vskip-1.0truein
\caption{Mass accretion rate $\dot M$ as a function of time for 
later epochs. The accretion rate is given in units of $M_\oplus$/Gyr
on the left vertical axis and in units of g/s on the right axis.
The upper curves show previous results for $\tau_1=10$ and different
collision locations, where $\xi_0=6$ (green), 8 (blue), and 10
(black). The cyan curves show results for collisions that take place
at time $t_1=10^8$ yr, with starting locations $\xi_0$ = 6, 8, 10, 12,
14, and 16 (left to right). All cases use starting mass $M_0=1M_\oplus$.
The horizontal dashed lines delineate the range of observed mass
accretion rates. The red curve shows the benchmark mass accretion rate
for the limit of maximally efficient collisions (from equation
[\ref{mdotgrind}]); for sufficiently long times, the WD luminosity
is too weak to expel even the smallest grains and the benchmark curve
is depicted as dotted. }
\label{fig:mdotlimit}
\end{figure}

{The solutions presented thus far are valid for a given distribution
of rock sizes, where Figures \ref{fig:sigma} -- \ref{fig:mdot} use
$dm/d\swig$ = {\sl constant} with $\smin=0.1$. This distribution is
expected for early collisions ($\tau_1=10$) when smaller rocks are
removed from the system by radiation pressure, but the distribution
does not change with time. For comparison, we can also consider the
case of much later collisions and more efficient collisional grinding.
Figure \ref{fig:mdotlimit} shows the mass accretion rates at later
epochs for these cases. The upper curves correspond to three of the
previous cases for starting time $\tau_1=10$ and $\xi_0$ = 6 (green),
8 (blue), and 10 (black). The cyan curves show the expected mass
accretion rates for collisions that take place at a later epoch
$t_1=10^8$ yr ($\tau_1=10^4$). For later collisions, we expect the
minimum rock size to be smaller (radiation pressure is less effective)
and these results use $\smin$ = 0.01. The six curves in the Figure
correspond to starting radii $\xi_0$ = 6, 8, 10, 12, 14, and 16 (from
left to right). As the starting radius grows, it takes longer from the
mass accretion flow to develop, and the rate is smaller (because the
larger rocks cannot move far enough inward in the available time).
The horizontal dashed lines (magenta) delineate the range of observed
(inferred) mass accretion rates. Finally, the red curve depicts the
benchmark mass accretion rate for the limit of maximally efficient
collisions (equation [\ref{mdotgrind}]).}

\section{Conclusion}
\label{sec:conclude}

This paper develops a new scenario for the production of debris disks
around white dwarfs. The basic features of the model can be summarized 
as follows:

Mass loss during the AGB phase of stellar evolution leads to
instability of planetary orbits. The planets must have semimajor axes
in the approximate range $a\approx3-10$ AU. Tighter orbits would cause
the planets to become engulfed by the star, whereas wider orbits would
lead to planet scattering instead of collisions. The resulting
collisions are sufficiently energetic to vaporize the planetary
material.  Since only a fraction of the planetary material
participates in the collision and vaporizes, the self-gravity of the
gas cloud is not strong enough to confine it.

The resulting cloud of silicate vapor (along with other elements)
expands and cools. The expanding cloud nucleates droplets of silicate
material, which subsequently hardens into rocks. The size of the
droplets is controlled by the Weber number and grows as long as the
cloud expands. The background density of the cloud decreases with
time, and thus allows for ever larger droplets to exist. Once the
cloud is too diffuse, however, merging droplets cannot find each other
and the size distribution freezes out. {The expanding cloud then 
becomes optically thin}, and the droplets cool into rocky bodies,
with a distribution of sizes, and characteristic size $s\sim1$ cm.
With their expected expansion velocities, the rocks formed in the flow
attain orbital eccentricities of order $e\sim0.1-0.2$, and thereby
spread over an annulus with width $w/r=\width/\xi_0\sim0.1-0.2$.

The maximum rock size is estimated to be of order 1 cm.  Since the
merging of droplets is efficient up to this size scale, one expects most
of the mass to reside in rocks near this limit, with some mass in
smaller bodies. {Sufficiently small rocks are removed by
radiation pressure from the star, where the size threshold depends
on the time of the planetary collisions.} 

The debris field produced by the collision starts as an annulus with
radius $r\sim several$ AU and width $w\sim r/10$. The collisions occur
while the star is still much brighter than its eventual white dwarf
luminosity. The luminosity that drives the PR effect is thus much
larger than is usually considered and is highly time dependent. We find
solutions for the orbital evolution of individual rocky bodies in this
scenario (Section \ref{sec:prdrag}) and for the surface density of the
entire debris disk (Section \ref{sec:evolve}).

Since the luminosity of the star decreases with time, the resulting
time dependence of the surface density, the mass of the debris disk,
and the accretion rate all depend on the time parameter
$\eta\sim\int L_\ast dt$ (where $\eta\sim\ln t$ for the fiducial case
with $L_\ast \propto 1/t$). In any case, these quantities are slowly
varying. Moreover, the time scales required to move rocky debris from
spatial {scales} of order several AU down to the white dwarf surface
are $\sim1$ Gyr. This `coincidence' --- that the moving time is
comparable to the age of the galactic disk --- is the same
`coincidence' that allows us to use the white dwarf luminosity
function to measure the age of the galactic components.

In physical units, the mass accretion rate starts large and decreases
down to values of ${\dot M}\sim10^9$ g/s at ages of $\sim1$ Gyr
(Figure \ref{fig:mdot}). The mass accretion rate can be larger for
earlier collisions (or larger starting mass) and can be smaller if the
collisions take place later or at larger distances from the star (or
with smaller starting mass). The expected late time accretion rates
are thus consistent with the observed estimates of
${\dot M}\sim10^6-10^{10}$ g/s.

{The preferred collision epoch at $\tau_1\sim10$
($t_1\sim10^{5}$~yr) is expected dynamically: Smooth, isotropic mass
loss drives multi-planet systems to destabilize on timescales up to
$\sim1$ Gyr. However, stellar mass loss is both episodic and
imperfectly isotropic. Mass ejection events produce recoil kicks with
$\Delta v=(M_{\rm ej}/M_\star)\,v_{\rm esc}$, which accumulate to
produce amplitudes $(\Delta{v})_T\sim0.5\;{\rm km\,s^{-1}}$ for
$M_{\rm ej}\sim10^{-4}\,M_\odot$, the value implied by white-dwarf
kinematics \citep{fuller2026}. These kicks act as a stochastic forcing
on any surviving planetary system \citep{batygin2026}. For planets
that escape engulfment and are delivered by adiabatic expansion to the
several-AU window relevant here, these kicks pump eccentricities to
$e\sim{\rm few}\times10^{-2}$ and collapse mutual Hill spacings well
beyond the adiabatic $(M_{\rm f}/M_{\rm i})^{1/3}$ reduction. These
changes in architecture drive most systems to orbit-crossing within
$\sim1$~Myr of white-dwarf formation, with about one third of the
ensuing instabilities leading to collisions rather than ejections.  A
dedicated calculation will be presented elsewhere; in the present
context, we note that $\tau_1\sim10$--$100$ is generic, and that our
results depend on $\tau_1$ only logarithmically. }
 
For completeness, note that the framework developed here can be
expanded in a number of directions. One can consider more detailed
models for the production of rocky bodies after the collision and make
an improved estimate for their size distribution. After the
condensation of the gas cloud into rocks, we have modeled the starting
state with a gaussian form, and considered the width of the annulus to
be $\width\sim\xi_0/10$ (corresponding to initial orbital
eccentricities $e_0\sim0.1$). Again, more detailed models can be
constructed.  In particular, multiple collisions can take place, which
would lead to multiple clouds of gas and rock that evolve according to
the scenario developed here. The detailed results of this scenario
depend on the luminosity evolution of the star, from its AGB phase to
the WD cooling phase. More detailed models (beyond that of equations
[\ref{lstart}] and [\ref{wdcool}] used here) are thus indicated. 

Because observed white dwarf enrichment events are common, any viable
explanation must involve physical processes that occur frequently.
The scenario presented here {has a number of features that are} 
consistent with this requirement: The planets required for collisions
are superearths -- the most common type of planet. The required
location for the collisions falls at several astronomical units, near
the ice-line of the original disk, and where planets have the highest
occurrence rate.  Moreover, collisions readily produce rocky debris
when they take place at non-zero impact parameter, the most common
type of collisions. With these properties, the collisions result in
vaporization, followed by nucleation and merging of droplets, and
produce rocky bodies of characteristic size $s\sim1$ cm. Bodies of
this size are also produced by coagulation, and by collisions of
larger rocks that are known to shatter. Moreover, they are large
enough to avoid expulsion from radiation pressure, and small enough to
allow for inward transport via PR drag on Gyr time scales. This
scenario thus {utilizes physical processes that commonly occur 
in post-main-sequence environments, as needed to explain} the observed
ubiquity of white dwarf pollution. 

\bigskip
$\,$ 
\centerline{\bf Acknowledgments}

We are grateful to Valeriya Kachmar and Siyi Xu for useful
discussions. We also thank the referee, Alexander Mustill, for a
comprehensive review that helped improve the paper. This work was
supported by the National Science Foundation (grant nos. 2508843 and
2408867), the University of Michigan, the California Institute of
Technology, the Leinweber Institute for Theoretical Physics (at
U. Michigan), and by the David and Lucile Packard Foundation.

\appendix
\section{Orbital Evolution of Rocks from the PR Effect:
  Radiation Pressure} 
\label{sec:orbradiation}

For completeness, this Appendix derives a more general description
for the orbital evolution of rocky bodies subjected to PR drag. 
If we include the outward force of radiation pressure, then we must
consider the two-dimensional problem. The radial equation of motion
takes the form
\be
{\ddot r} - r {\dot\theta}^2 = - {GM\over r^2} (1 - \beta) 
\qquad {\rm where} \qquad 
\beta\equiv {3L_\ast \over 16\pi c\rho GMs} \,.
\label{betadef} 
\ee
Note that in general $L_\ast$ is a function of time so that
$\beta=\beta(t)$. The azimuthal equation of motion has the form
\be
{dJ \over dt} = - \beta {GM\over c} {\dot\theta} \,,
\qquad {\rm where} \qquad 
J = r^2 {\dot\theta} \,.
\ee
Next we consider nearly circular orbits, where this approximation
requires that the time scale for the orbit to change is much longer
than the orbit time (see Appendix \ref{sec:orbeccentric}). 
As a result, we can write
\be
v_\theta = r
{\dot\theta} = \left[ {GM(1-\beta)\over r}\right]^{1/2} \,,
\ee
so that the angular momentum varies according to 
\be
{dJ \over dt} = - \beta {GM\over c}
{\sqrt{GM(1-\beta)}\over r^{3/2}} \,.
\ee
In this approximation, the acceleration term in the radial equation of
motion is much less than the other two terms:
\be
|{\ddot r}| \ll | r {\dot\theta}^2 | \sim
{GM\over r^2} (1-\beta) \,.
\label{ordering} 
\ee
In this limit, the angular momentum takes the form
$J = \sqrt{GM(1-\beta)r}$, so that  
\be
{dJ\over dt} = {1\over2} \sqrt{GM(1-\beta)} {{\dot r}\over\sqrt{r}}
- {\dot\beta} \sqrt{GMr} {1 \over 2 \sqrt{1-\beta}} \,. 
\ee
If we use this expression in the azimuthal equation of motion,
we find
\be
{ {\dot r} \over r} = { {\dot\beta} \over 1-\beta}
   - {2GM\over c} {\beta\over r^2} \,. 
\ee
To solve this equation, we first define
\be
f \equiv r (1-\beta) \,, 
\ee
so that the equation of motion becomes
\be
f {df\over dt} = - {2GM\over c}
\beta (1-\beta)^2 \,, 
\ee
Integrating from some starting time $t_0$ to final time $t_f$,
we find
\be
f^2 (t_f) - f^2 (t_0) = - {4GM\over c} \int_{t_0}^{t_f} dt
\beta (1-\beta)^2 \,. 
\ee
To move forward we need to specify the time dependence of the
stellar luminosity and hence the time dependence of $\beta$.
Following the approximations that $L_\ast\sim1/t$, we can take
\be
\beta = \beta_0 / \tau \qquad {\rm where} \qquad
\tau = t / t_0 \,. 
\ee
Our solution then becomes 
\be
f^2 (t_f) - f^2 (t_0) = - {4GM \beta_0 t_0 \over c}
\int_1^{\tau_f} {d\tau \over \tau} \left( 1 - \beta_0/\tau \right)^2\,,
\ee
which evaluates to the form 
\be
f^2 (t_f) - f^2 (t_0) = - {4GM \beta_0 t_0 \over c}
\left[ \ln \tau_f - 2 \beta_0 + \beta_0^2 /2 \right] \,,
\ee
where $\tau_f\gg1$.  Note that if the problem starts with
$\beta_0\ll1$, then the solution reduces to the form 
\be
r_0^2 - r_f^2 = {4GM \beta_0 t_0 \over c}
\ln \tau_f \,,
\ee
as derived previously in the main text. Notice also that for this
solution, the ordering of equation (\ref{ordering}) is in fact
satisfied, i.e., the radial acceleration is small.

\section{Orbital Evolution of Rocks from the PR Effect:
  Nonzero Eccentricity} 
\label{sec:orbeccentric}

The main text focuses on the case of circular orbits for the rocky
debris. This appendix considers the more general case of eccentric
orbits and quantifies the degree of approximation made previously.
The correction to the radial motion due to nonzero eccentricity is
estimated to be $\sim3\%$ at the start of the evolution, and decreases
more rapdily than the semimajor axis.  We start by defining the
parameter
\be
\mu = GM(1-\beta),
\ee
where $\beta$ is given by equation (\ref{betadef}). 
To first order in $v/c$, the Poynting--Robertson acceleration
\citep{burns1979} can be written in the form  
\be
\mathbf a_{\rm PR} = -\frac{\beta\mu}{cr^2}
\left(
2v_r\hat{r} + v_\theta\hat{\theta} 
\right).
\ee
Throughout this discussion we assume that the orbit time is much
smaller than the evolution time, $P_{\rm orb}\ll t_{\rm evol}$, so
that $\beta$ can be considered constant over one orbit.

The specific orbital energy is given by 
\be
E = \frac{v^2}{2} - \frac{\mu}{r}\,,
\ee
and the corresponding rate of energy loss due to PR drag becomes 
\be
{\dot E} = \mathbf v\cdot\mathbf a_{\rm PR} =
-\frac{\beta\mu}{cr^2} \left( 2v_r^2+v_\theta^2 \right). 
\ee
Next we invoke the standard Keplerian relations 
\be
E=-\frac{\mu}{2a},
\qquad
h^2=\mu a(1-e^2),
\qquad 
r = \frac{a(1-e^2)} {1+e\cos\theta},
\label{keplerone} 
\ee
with the velocity fields 
\be
v_r = \sqrt{\frac{\mu}{a(1-e^2)}} \,e\sin\theta 
\qquad {\rm and} \qquad 
v_\theta = \sqrt{\frac{\mu}{a(1-e^2)}} (1+e\cos\theta). 
\ee
Because the evolution time is long compared to the orbital
period, we perform orbit averaging. The increment of time
is related to the angle $\theta$ according to
\be
dt=\frac{r^2}{h}d\theta\,.
\ee
The orbit-averaged rate of energy dissipation and the corresponding
change in semi-major axis are given by 
\be
\langle {\dot E}\rangle = -
\frac{\beta\mu^2} {ca^3(1-e^2)^{3/2}} \left(1+\frac32e^2\right) 
\qquad {\rm and} \qquad  
\langle{\dot a}\rangle = - \frac{2\beta\mu}{ca}
\frac{1+\frac32e^2}{(1-e^2)^{3/2}}.
\ee
Note that the latter equation reduces to the circular case
from equation (\ref{motionone}) in the appropriate limit.
Moreover, the leading order correction due to nonzero eccentricity
introduces a factor of $(1 + 3e^2 + {\cal O}[e^4])$. Because the
starting eccentricity $e_0\sim0.1$ (see main text), the correction
in only $\sim3\%$.  

Next we consider the evolution of angular momentum and
orbital eccentricity. The time rate of change of angular
momentum due to PR torque is given by 
\be
\frac{\dot h}{h} = -\frac{\beta\mu}{cr^2} 
\qquad \Rightarrow \qquad 
\left\langle \frac{\dot h}{h} \right\rangle = - 
\frac{\beta\mu}{ca^2\sqrt{1-e^2}}\,,
\ee
where the final expression gives the orbit-averaged
quantity. The orbital angular momentum and eccentricity
are related through equation (\ref{keplerone}) so that 
\be
2\frac{\dot h}{h} =
\frac{\dot a}{a} - \frac{2e\dot e}{1-e^2}
\ee
Solving for the eccentricity evolution, one finds 
\be
\langle {\dot e} \rangle = - 
\frac{5\beta\mu}{2ca^2} \frac{e}{\sqrt{1-e^2}}. 
\ee
Next we combine the time evolution equations for semimajor
axis and eccentricity to find 
\be
{5\over4} {da \over a} = {de \over e} 
\frac{1+\frac32e^2}{1-e^2}.
\ee 
This equation can be integrated to obtain
\be
e = e_0 \left({a\over a_0}\right)^{5/4}
\left( {1-e^2\over1-e_0^2}\right)^{5/4} \,. 
\ee
The eccentricity thus decreases more rapidly than the
semimajor axis, with the leading order behavior given by
$e \propto a^{5/4}$.  

\section{Half-Life of the Debris Disk}
\label{sec:halflife}

We can define the half-life of the disk to be the time required for
half of the disk mass to be accreted onto the central star.  This
Appendix shows that for a given starting location $\xi_0$, which
represents where the original planetary collision occurs, the disk
half-life has the form 
\be
{\eta_{1/2} \over \xi_0^2} = constant \approx 0.274 \,.
\label{etahalf}    
\ee
The mass of the disk at any time, specified by the parameter $\eta$,
is given by
\be
M(\eta) = 2\pi r_0^2 \Sigma_0 \int_{\smin}^1 {d\swig\over1-\smin} 
\int_0^\infty \xi d\xi \exp\left[ - \left( 
\sqrt{\xi^2 + 2\eta/\swig} - \xi_0 \right)^2 /\width^2 \right] \,, 
\ee 
Now we change variables to $z=\xi/\xi_0$, write the expression in
terms of the starting mass $M_0$, and use the fact that the width of
the starting distribution $\width=\xi_0/10$. The resulting expression
for the mass takes the form
\be
M(\eta) = {10 M_0 \over \sqrt{\pi}}
\int_{\smin}^1 {d\swig\over1-\smin} 
\int_0^\infty zdz \exp\left[ - 100 \left(
\sqrt{z^2 + h/\swig} - 1 \right)^2 \right] \,, 
\ee
where we have defined a scaled time variable
\be
h \equiv 2\eta / \xi_0^2 \,.
\ee
The half-life is given by the value of $h$ such that 
\be
\int_{\smin}^1 {d\swig\over1-\smin} 
\int_0^\infty zdz \exp\left[ - 100 \left(
\sqrt{z^2 + h/\swig} - 1 \right)^2 \right] = {\sqrt{\pi} \over 20} \,.
\ee
The left-hand-side of the equation is a function of the variable $h$
only (for a given value of minimum value of $\swig$, taken to be
$\smin$ = 0.1 here).  The resulting value of $h$ (that satisfies the
equality) is thus independent of the starting location $\xi_0$.
Numerical evaluation of the integral indicates that the required value
of $h\approx0.547$, which in turn results in the scaling law of
equation (\ref{etahalf}). For completeness, one can define $\eta_{90}$
to be the time parameter when 90\% of the disk is accreted, which
results in the analogous scaling law $\eta_{90}\approx0.468\xi_0^2$.

\section{Upper Limit on Rock Transfer from Global Energy Considerations}
\label{sec:upperlimit}

As a consistency check, we can calculate the amount of energy required
for a star to move a rocky body. This consideration can then be used
to place an upper limit on the amount of mass that can be transferred.
Although the body is falling deeper into the gravitational potential
well of the star, and doesn't need to absorb energy to move inward,
the process is driven by starlight impinging on the rock and being
re-radiated (to transfer angular momentum). The energy required to
drive this process is absorbed by the rock at the rate 
\be
{\dot E} = \pi s^2 {L_\ast \over 4\pi r^2} \,. 
\ee
Given the rate at which the rock moves inward,
\be
{\dot r} = - {3L_\ast \over 8\pi\rho sc^2 r} \,, 
\ee
we find the energy required to move the rock inward must
obey the differential equation 
\be
\Big|{dE \over dr}\Big| = \pi s^2 {L_\ast \over 4\pi r^2}
{8\pi\rho sc^2 r \over 3L_\ast} = {1\over2} {mc^2 \over r} \,. 
\ee
After integrating, the energy increment needed to move a
rock of mass $m$ from starting location $r_i$ to final
location $r_f$ is given by
\be
\Delta E = mc^2 {\ln(r_i/r_f)\over2} \approx 5 mc^2 \,,
\ee
where we have taken $r_i/r_f\sim3\times10^4$ in the final approximate
equality.  The total mass $M_T$ of rock that can be moved is given by
the total energy radiated during the cooling phase, namely $E_{\rm WD}$,
so that 
\be
M_T = {2 E_{\rm WD} \over c^2 \ln(r_i/r_f)} \approx 7.2
M_\oplus\,. 
\ee
In practice, only a fraction $f_\gamma$ of the stellar photons travel
close enough to the orbital plane of the rocks to be intercepted.
Here we assume that the height of the rocky disk is roughly comparable
to its radial extent, with an aspect ratio of 1:10, so that
$f_\gamma\approx0.1$.  As a result, a working upper limit for the
amount of rock transfered is an order of magnitude smaller. The upper
limit on the total mass becomes 
\be
M_T \lta {2 f_\gamma E_{\rm WD} \over c^2 \ln(r_i/r_f)} 
\lta 1 M_\oplus\,. 
\ee

\end{document}